\documentclass[lettersize,journal]{IEEEtran}
\usepackage{amsmath,amsfonts}
\usepackage{algorithmic}
\usepackage{algorithm}
\usepackage{array}
\usepackage[caption=false,font=normalsize,labelfont=sf,textfont=sf]{subfig}
\usepackage{textcomp}
\usepackage{stfloats}
\usepackage{url}
\usepackage{verbatim}
\usepackage{graphicx}
\usepackage{cite}
\usepackage{amsmath}
\usepackage{siunitx}
\usepackage{bm} 
\usepackage{subcaption}
\newcommand{\ie}{{\it i.e.}}
\newcommand{\eg}{{\it e.g.}}
\newcommand{\etal}{{\it et al.}}
\usepackage{booktabs}
\usepackage{color}
\usepackage{soul}
\begin{document}

\title{360CityGML: Realistic and Interactive Urban Visualization System Integrating CityGML Model and 360° Videos}

\author{Tatsuro Banno, Mizuki Takenawa, Leslie W{\"o}hler, Satoshi Ikehata, Kiyoharu Aizawa

\thanks{This work was partially supported by JSPS KAKENHI 21H03460 and CISTI SIP.}%
\thanks{Tatsuro Banno, Mizuki Takenawa, and Leslie W{\"o}hler are with the University of Tokyo, Tokyo, Japan (e-mail: banno@hal.t.u-tokyo.ac.jp; takenawa@hal.t.u-tokyo.ac.jp; woehler@hal.t.u-tokyo.ac.jp).}
\thanks{Satoshi Ikehata is with National Institute of Informatics, Tokyo, Japan (sikehata@nii.ac.jp).}
\thanks{Kiyoharu Aizawa is with the University of Tokyo, Tokyo, Japan (e-mail:aizawa@hal.t.u-tokyo.ac.jp).}
}

\maketitle

\begin{abstract}
We introduce a novel urban visualization system that integrates 3D urban model (CityGML) and 360° walkthrough videos. By aligning the videos with the model and dynamically projecting relevant video frames onto the geometries, our system creates photorealistic urban visualizations, allowing users to intuitively interpret geospatial data from a pedestrian view.
\end{abstract}


\section{Introduction}
\IEEEPARstart{C}{}ityGML (City Geography Markup Language) is an open, standardized data format for urban 3D models. CityGML integrates various types of geographic data, such as geometry, building attributes and disaster risk information, into a unified geospatial framework~\cite{groger2012citygml}, making it a crucial tool for comprehensive analyses of urban scenes~\cite{PLATEAU, opencitymodel, dollner2006virtual}.

Various visualization techniques have been proposed to effectively analyze urban 3D data~\cite{miranda2024state}. Some employ a distant bird's-eye perspective for visualization~\cite{yao20183dcitydb, johnston2001using, cornel2019interactive, miranda2018shadow}, which is useful for macro-level analysis. On the other hand, to analyze the ground-level experience of urban environments, several works focus on pedestrian-view visualizations~\cite{bartosh2019immersive, zhang2021urbanvr, boorboor2024submerse}, aiming to create more immersive urban experiences.

\begin{figure}[t]
    \centering
    \includegraphics[width=\linewidth]{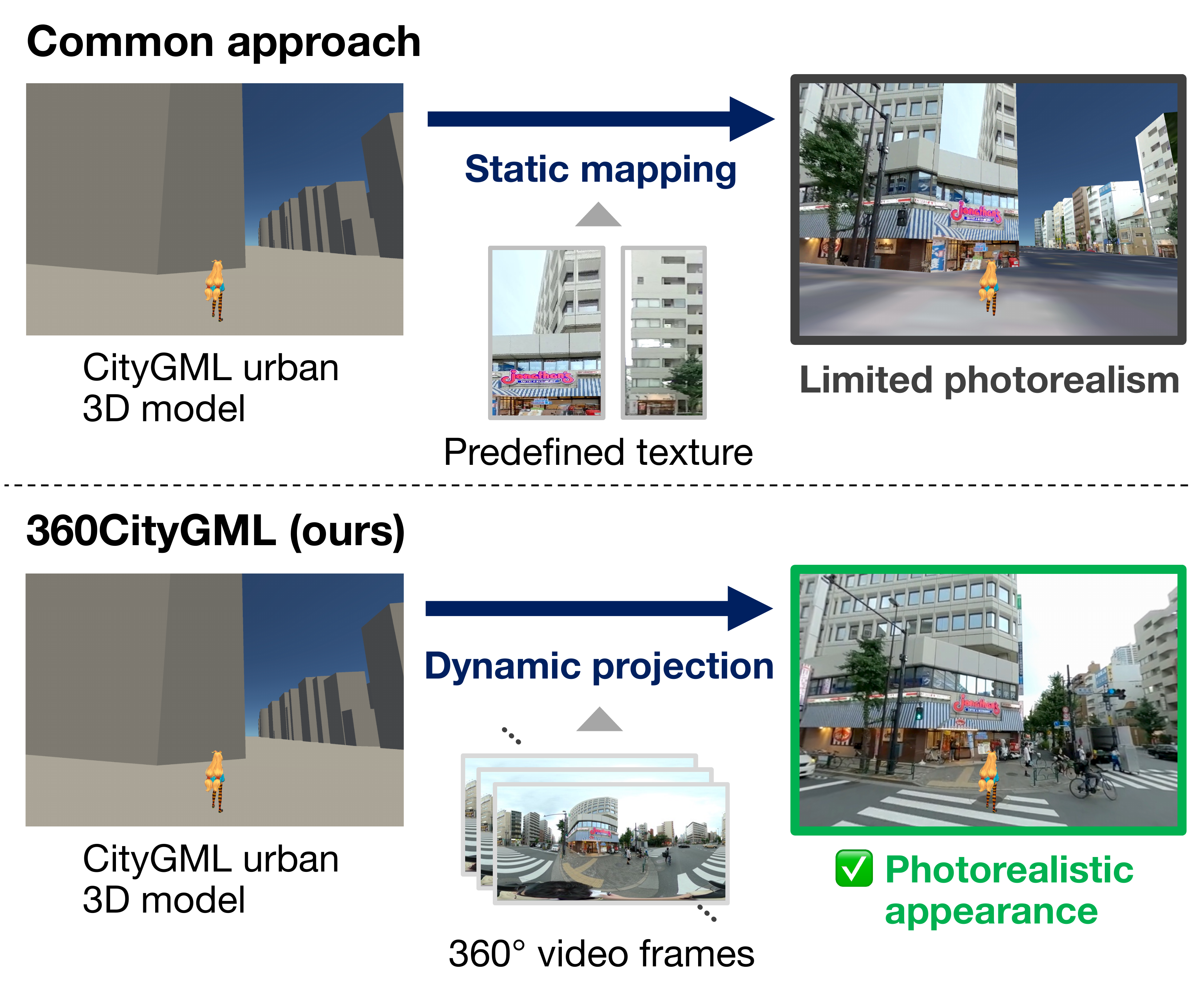}
    \caption{Comparison between the conventional static texture mapping of each face~\cite{boorboor2024submerse}~(top), and our dynamic texture projection of 360° video frames~(bottom). Our approach faithfully reproduces the real-world appearance to a much greater extent.}
    \label{fig:texture_compare}
\end{figure}

To enhance the sense of immersion in urban experiences, photorealism is essential. A basic approach to enhancing photorealism involves using meticulously crafted, high-quality 3D meshes~\cite{zhang2021urbanvr} with static texture mapping~\cite{boorboor2024submerse}. However, most CityGML datasets in public offer highly simplified geometry (\ie, Level of Detail 1, or LOD1)~\cite{opendatasets}, typically represented as basic box-shaped structures. Applying texture mapping to such simple models often results in noticeable inconsistencies between the simplistic geometry and the textures, significantly diminishing realism (Fig.\ref{fig:texture_compare}-top).

Meanwhile, researchers have recently developed systems for virtual exploration of urban environments using 360° walkthrough videos recorded along streets in the target area~\cite{sugimoto2020urban, sugimoto2020building, takenawa2023360rvw, takenawa2025building}. Unlike regular videos, 360° videos capture the entire surroundings, allowing users to freely change their viewing direction for a more immersive experience. Compared to simply visualizing texture-mapped 3D models, these systems provide a richer visual experience by directly displaying the original videos from the cameraman's perspective. However, the absence of 3D information in the scene limits their applicability for visualizing 3D geographical data for the viewer such as visualizing 3D flood data overlaid to the rendered view.

In this paper, we propose \emph{360CityGML}, a realistic urban 3D visualization system that integrates the strengths of CityGML models and 360° walkthrough videos. This system addresses the loss of photorealism caused by na\"ive static texture mapping on CityGML models by dynamically utilizing the photorealism of 360° video footage. Specifically, it projects 360° video frames onto the CityGML geometry as textures, aligning them with the original camera positions of the 360° video footage during visualization, thereby preventing users from noticing artifacts caused by the simplified geometry. To reduce the perception of viewpoint constraints, the system dynamically selects the most appropriate 360° video frame and its spherical orientation based on the avatar's location and movement direction within the CityGML model. This approach enables users to virtually walk through the environment using an avatar, while maintaining seamless and photorealistic visualization (Fig.~\ref{fig:texture_compare}-bottom).

To integrate CityGML models and 360° walkthrough videos, precise alignment between the two modalities is essential. Since frame-by-frame alignment between 3D geometry and 360° videos is computationally prohibitive~\cite{taneja2012registration}, we propose a street-level alignment approach. In this approach, the local camera trajectory reconstructed by Visual SLAM~\cite{sumikura2019openvslam} is aligned with the 3D model by minimizing the differences between semantic maps derived from the 3D geometry and a few sampled key frames.

We demonstrate the system’s effectiveness in the urban area of Akihabara, Japan, showcasing its application in flood risk visualization and daylight visualization. In addition, we conducted a user study focusing on flood risk scenarios to evaluate how photorealistic 360° videos enhance users' comprehension of geospatial data through immersive experiences.

\section{Related Work}

\subsection{3D urban visualization systems from a pedestrian view}

We focus on pedestrian-view visualization systems~\cite{bartosh2019immersive, zhang2021urbanvr, boorboor2024submerse}, which are central to our approach (see~\cite{miranda2024state} for a broader survey). Bartosh~\etal~\cite{bartosh2019immersive} developed a VR system visualizing urban data like transit and noise, but with simple, non-photorealistic models. UrbanVR~\cite{zhang2021urbanvr} enables immersive pedestrian navigation with LOD3 geometries and high-resolution textures, though its reliance on LOD3 limits its general usability. Submerse~\cite{boorboor2024submerse} visualizes flood scenarios using box-shaped 3D models of New York City and high-quality textures, but static texture mapping reduces realism due to mismatches with geometry.

In contrast, we propose a novel system that combines CityGML models with 360° video, allowing highly photorealistic pedestrian-view visualization by directly using captured video for rendering.

\subsection{Visual enhancement of 3D urban model with textures}

Enhancing the visual realism of urban 3D models with simplified geometry through texture mapping remains a long-standing challenge. A common approach maps predefined static textures onto model surfaces~\cite{blender, boorboor2024submerse}. However, for oversimplified CityGML LOD1 models, such naïve mapping often introduces significant discrepancies between the geometry and the textures, severely limiting photorealism.

To overcome this limitation, several studies have explored projective texture mapping of street view images~\cite{bredif2013image,bredif2014projective, du2019geollery,park2021instant}. However, these methods still suffer from noticeable discrepancies between geometry and textures, as they simply map the nearest-neighbor image~\cite{du2019geollery} or blend multiple nearby images onto the geometry~\cite{bredif2013image, bredif2014projective, park2021instant}, resulting in visible artifacts for viewpoints outside the captured positions. To address these issues, we propose dynamically projecting 360° video frames onto the CityGML model while constraining the camera view to the original 360° video trajectory. This approach ensures a photorealistic experience and effectively eliminates discrepancies between geometry and textures.

\subsection{Application of 360° videos for virtual exploration}
360° videos capture the entire surroundings simultaneously, enabling immersive experiences where users can freely explore by changing their viewing direction. Building on this concept, Movie Map~\cite{sugimoto2020urban, sugimoto2020building} was developed using 360° walkthrough videos of multiple urban streets for virual exploration of the area. By leveraging coordinates and camera trajectories obtained via Visual SLAM~\cite{sumikura2019openvslam}, the system automatically detects intersections along routes, allowing seamless transitions between video segments for continuous exploration. These segments have also been used to create virtual worlds navigable by avatars~\cite{takenawa2023360rvw, takenawa2025building}.

However, conventional methods are limited to video-based urban visualization, as they lack the scene's 3D geometric information. In contrast, our approach spatially aligns 360° video with CityGML-based urban data, enabling coherent and consistent 3D urban visualization for applications such as flood risk and daylight analysis.

\subsection{Urban modeling via novel view synthesis methods}
Recent advances in view synthesis, such as NeRF~\cite{mildenhall2020nerf} and 3D Gaussian Splatting~\cite{kerbl3Dgaussians}, have enabled photorealistic 3D reconstruction of urban environments from cityscape images~\cite{liu2024citygaussian, otonari2024entity}. However, these methods are not specifically designed for urban visualization and require both alignment with geospatial data and extensive training. In contrast, we propose a simpler and more practical approach: projecting 360° video directly onto CityGML models. This method enables effective urban visualization without large-scale training. Moreover, unlike NeRF and 3D Gaussian Splatting, which generate only static views, our approach can prroduce dynamic scenes with moving people and objects. 

\section{System Overview}
\begin{figure*}[t]
    \centering
    \includegraphics[width=.94\linewidth]{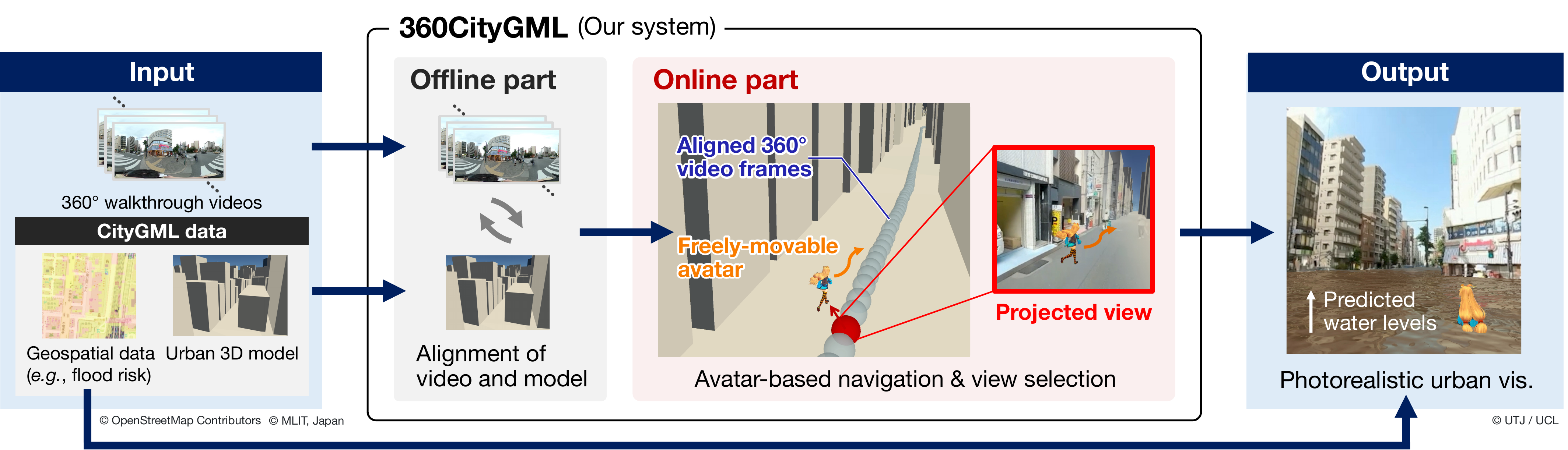}
    \caption{An overview of \emph{360CityGML}. Our system first aligns each frame of the 360° video with the CityGML model in an offline process. In the online system, a user can virtually walk through the environment by controlling their avatar. Based on the avatar's location and its movement direction, the system automatically selects the most suitable 360° video frame and its spherical orientation as the background. This frame is then directly projected onto the 3D urban model as a texture, delivering a photorealistic visual experience. Finally, the textured models are overlaid with geospatial data, offering highly photorealistic and 3D-aware urban visualizations.}
    \label{fig:overview}
\end{figure*}

We propose \emph{360CityGML}, a system designed to generate photorealistic 3D urban visualizations that effectively highlight geospatial data while allowing users to interactively move around the environment using avatars. An overview of the system is shown in Fig.~\ref{fig:overview}.

Our system uses two main inputs: \emph{1) 360° walkthrough videos of the target area.} In our approach, multiple walkthrough videos serve as input, each capturing a single street segment. The start and end points of each video are manually annotated by referencing their locations on a map during data acquisition. This street-by-street recording and annotation strategy enables efficient data collection and scales well to large urban areas, as demonstrated in previous virtual exploration systems.~\cite{sugimoto2020urban, sugimoto2020building, takenawa2023360rvw, takenawa2025building}. \emph{2) CityGML data for the target area.} This consists of geospatial data (\eg, flood risk) and urban 3D model (\eg, buildings and terrain). Our system specifically handles simplified, untextured geometric model (\ie, LOD1), which are commonly used in many CityGML datasets~\cite{opendatasets}.

Based on these inputs, our system first aligns each frame of the 360° video with the CityGML model in an offline process. In the online system, users can virtually walk through the environment by controlling their avatars. Based on the avatar’s location and movement direction within the CityGML model, the system dynamically selects the most suitable 360° video frame and its spherical orientation as the background. This frame is directly projected onto the 3D urban model as a texture, providing a photorealistic visual experience without noticeable discrepancies. Finally, the textured models are overlaid with geospatial data, offering highly realistic and 3D-aware urban visualizations in a real-world context.

To implement such a system, two primary requirements must be addressed. First, \emph{each video frame must be precisely aligned with the global CityGML coordinate system.} This alignment is crucial for the accurate visualization of geospatial data. However, since capturing large city areas leads to long videos consisting of a large number of frames, the prior frame-by-frame alignment method~\cite{taneja2012registration} is computationally impractical. To overcome this challenge, we propose a street-level alignment approach. In this method, the local camera trajectory of each street video is first estimated using Visual SLAM~\cite{sumikura2019openvslam}, and afterwards the entire trajectory is registered to the global coordinate system based on a subset of frames.

Second, \emph{the system must select an appropriate viewpoint and view orientation to seamlessly follow the avatar's movement and maintain photorealistic rendering quality}. This is essential for enabling users to smoothly walk through the environment with a freely movable avatar. Unlike prior works that introduce avatars in 360° images~\cite{chalmers2024avatar360} or videos~\cite{takenawa2023360rvw, takenawa2025building}, where the avatar moves in local camera-centered coordinates, our system must handle avatar movement within a global coordinate system while selecting viewpoints only from the camera's trajectory. To address this unique challenge, we propose a practical view-selection strategy: choosing viewpoints that maintain a consistent distance from the avatar along its forward trajectory and adjusting the view orientation within the 360° frame to smoothly face the avatar. 

The next section outlines the technical details of these approaches.

\section{Technical Detail}
\subsection{Alignment of CityGML urban 3D model and 360° video frames (Offline Part):}
\label{alignment}

The goal of this part is to precisely align the large-scale 360° walkthrough video frames in the 3D coordinate system of the CityGML model, where each coordinate corresponds to latitude, longitude, and height of the real world. This alignment task can be formulated as estimating the camera's position ($\bm{x} \in \mathbb{R}^3$) and orientation ($\bm{\theta} \in \mathbb{R}^3$) for all 360\textdegree~video frames in the coordinate system. Since prior frame-by-frame alignment~\cite{taneja2012registration}, where $\bm{x}$ and $\bm{\theta}$ were optimized independently for each frame, is computationally intractable for the urban-scale videos, we instead estimate the rigid transformation of street-level camera trajectories obtained by Visual SLAM~\cite{sumikura2019openvslam} as detailed below.
\\

\begin{figure}[t]
    \centering
    \includegraphics[width=\linewidth]{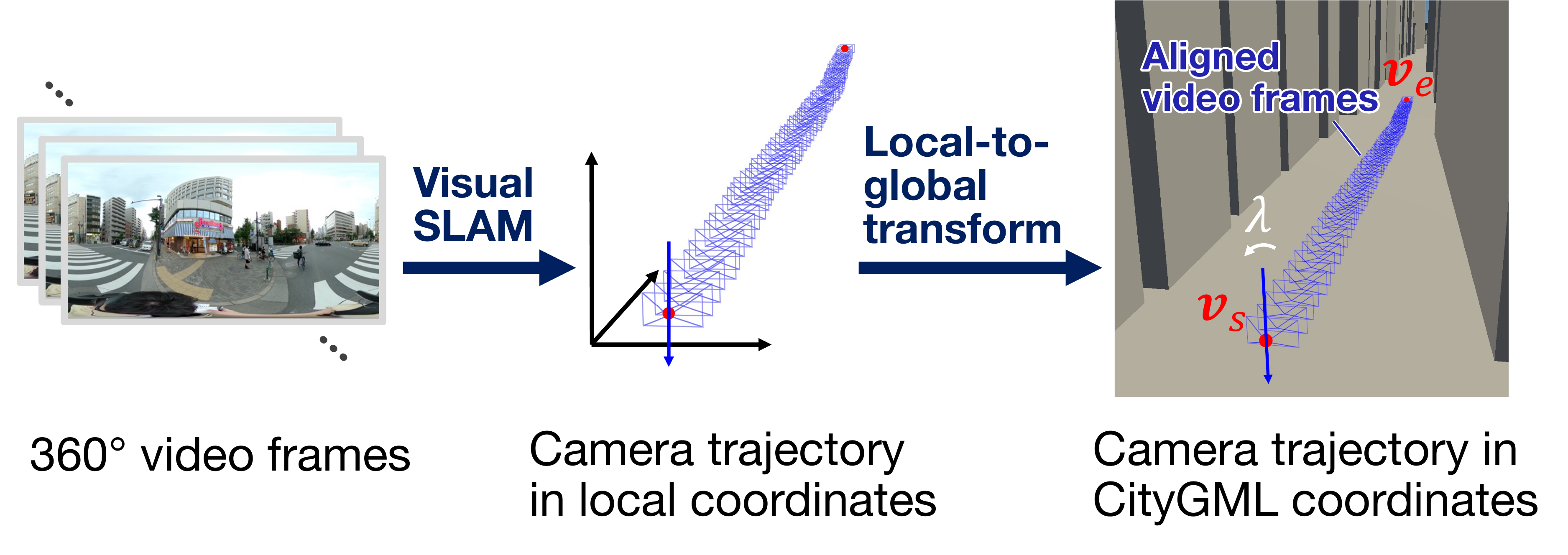}
    \caption{An overview of the formulation of our street-level alignment. First, we apply Visual SLAM~\cite{sumikura2019openvslam} to each 360° video to obtain the camera trajectory in local coordinates systems. Next, we perform a local-to-global transformation using the position of the camera's starting point $\bm{v}_s$, the position of the end point $\bm{v}_e$, and the rotation angle $\lambda$ around the gravity axis. This transformation aligns the video frames with the global CityGML coordinate system.}
    \label{fig:formulation}
\end{figure}

\noindent\emph{Formulation of street-level alignment:}
An overview of the formulation of our street-level alignment is shown in Fig.~\ref{fig:formulation}. We first apply visual SLAM~\cite{sumikura2019openvslam} to 360° walkthrough videos individually and obtain the cameras' local trajectories, which provide the position ($\bm{x}_\mathrm{local} \in \mathbb{R}^3$) and orientation ($\bm{\theta}_\mathrm{local} \in \mathbb{R}^3$) of each frame in their camera-specific coordinate systems. Then, we register the individual local trajectories to the global coordinate system.

To perform this registration, we compute the local-to-global transformation matrix for each trajectory that converts the camera-specific coordinate systems into the global CityGML coordinate system. 
This transformation matrix has seven degrees of freedom: three for translation, three for rotation, and one for scale. To fully define this matrix, we impose seven constraints, including the position of the camera's starting point $\bm{v}_s = (x_s, y_s, z_s)$, the position of the end point $\bm{v}_e = (x_e, y_e, z_e)$, and the rotation angle $\lambda$ around the gravity axis. By optimizing these parameters --- $\bm{v}_s$, $\bm{v}_e$, and $\lambda$ --- for each trajectory, we ensure precise alignment between the video frames and the CityGML model.
\\

\noindent\emph{Parameter optimization:}
We first initialize all parameters with initial values. Specifically, $x_s$, $y_s$, $x_e$, and $y_e$ are set from the manually annotated latitude and longitude of the starting and ending points. $z_s$ and $z_e$ are obtained by adding the camera height (2 meters in our experiment) to the corresponding CityGML terrain elevations, and $\lambda$ is initialized to 0° based on the gravity direction estimated by Visual SLAM.

Starting from these initial values, we optimize each parameter by minimizing the misalignment between the video frames and the CityGML model. To achieve this, we design a new objective function that effectively quantifies the misalignment by extending the building-region based approach used by frame-by-frame alignment~\cite{taneja2012registration} to the trajectory level.

\begin{figure}[t]
    \centering
    \includegraphics[width=\linewidth]{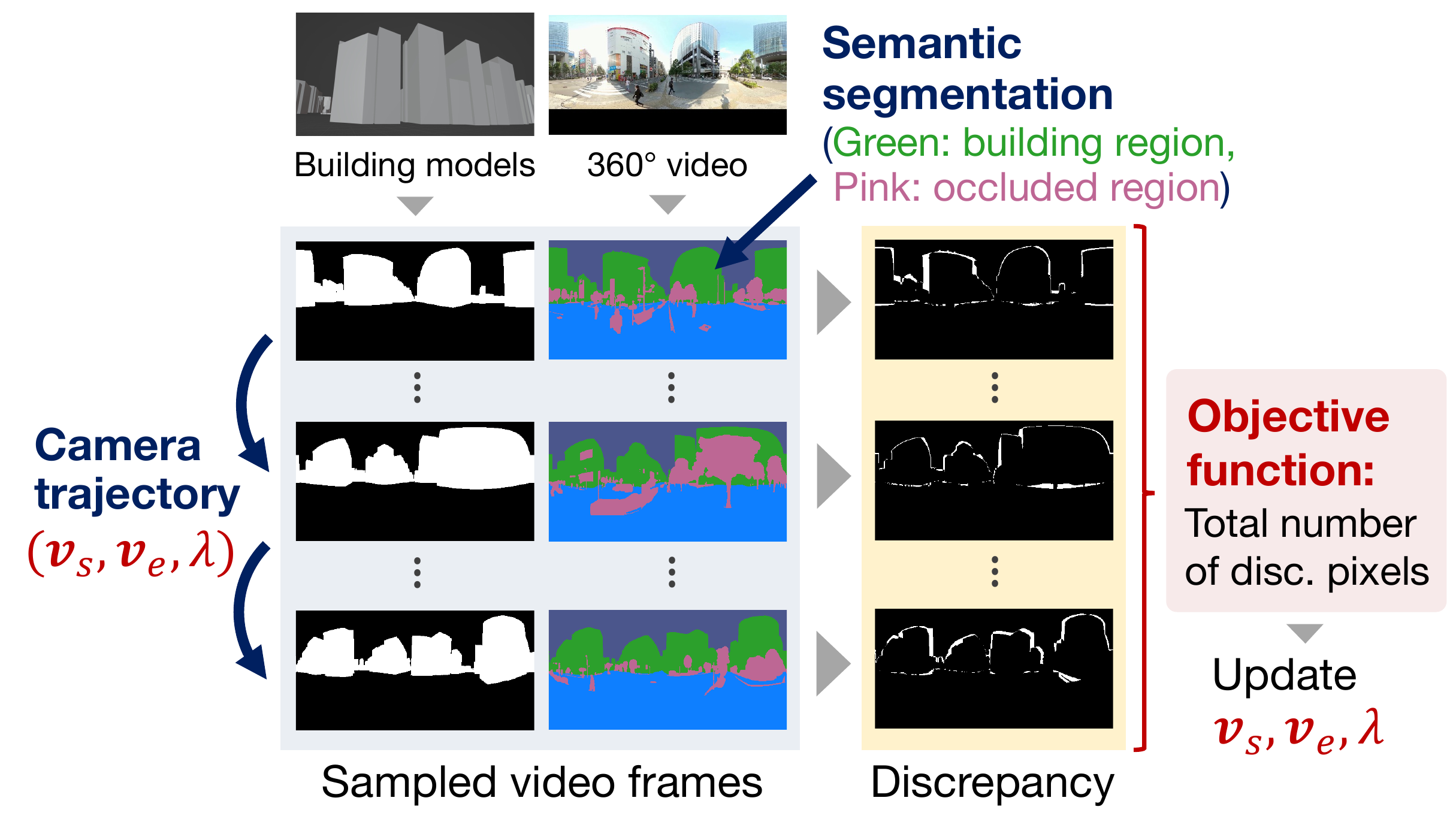}
    \caption{An overview of our objective function. For a sampled subset of video frame, we calculate the total number of discrepant pixels by comparing the building regions in the model projection to those in the video frames extracted by semantic segmentation~\cite{cheng2022masked}. We then optimize the parameters ($\bm{v}_s$, $\bm{v}_e$, $\lambda$) to minimize this discrepancy.}
    \label{fig:objective}
\end{figure}

Fig.~\ref{fig:objective} shows an overview of our objective function. To reduce computation, we sample $N = 30$ frames from the video at regular intervals and compute the objective function using only these frames. For each sampled frame, we evaluate discrepancies between the video frame and the CityGML model by focusing on building regions~\cite{taneja2012registration}. Specifically, we first extract building regions from the video frame using a semantic segmentation model~\cite{cheng2022masked} trained on a city landscape dataset~\cite{neuhold2017mapillary}. Next, using the estimated trajectories determined by the current parameters $\bm{v}_s$, $\bm{v}_e$, and $\lambda$, we project the CityGML building models onto a 360° view from the estimated camera pose of each frame. We then compare the building regions from the video frame and the model projection to evaluate pixel-wise discrepancy between them. Finally, we sum the number of discrepant pixels across all sampled frames, and use this value as the objective function. The parameters ($\bm{v}_s$, $\bm{v}_e$, and $\lambda$) are then iteratively updated using the Covariance Matrix Adaptation Evolution Strategy (CMA-ES)~\cite{hansen2001completely} to minimize the objective function. This update process is repeated 700 times, which are basically sufficient for the convergence.

\subsection{Dynamic view-selection of 360° video frames for avatar-based navigation (Online Part):}

In the online process, a user can virtually walk through the environment by moving their avatar. Based on the avatar’s location and movement direction, the system selects the most suitable 360° video frame and its spherical orientation for texture projection onto the CityGML urban 3D model. Notably, to mitigate discrepancies between geometry and textures, the viewpoints are restricted to the video camera’s trajectory. In this setting, we propose a practical strategy for selecting the viewpoint and view orientation to smoothly follow the avatar's movement and convey the sense of exploration.
\\

\begin{figure}[t]
    \centering
  \subfloat[\label{1a}]{%
       \includegraphics[width=0.48\linewidth]{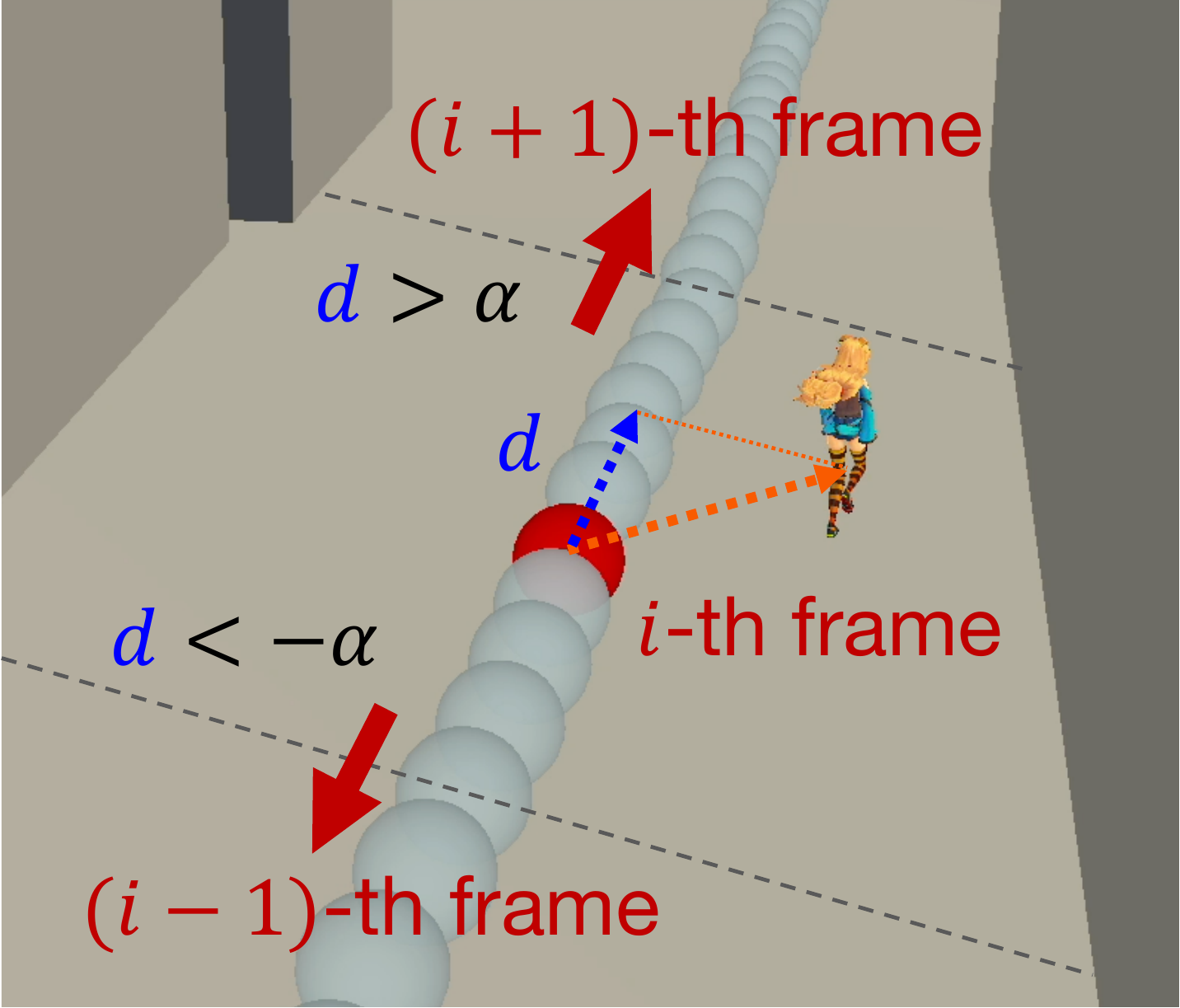}}
\hspace{0.01\linewidth}
  \subfloat[\label{1b}]{%
        \includegraphics[width=0.48\linewidth]{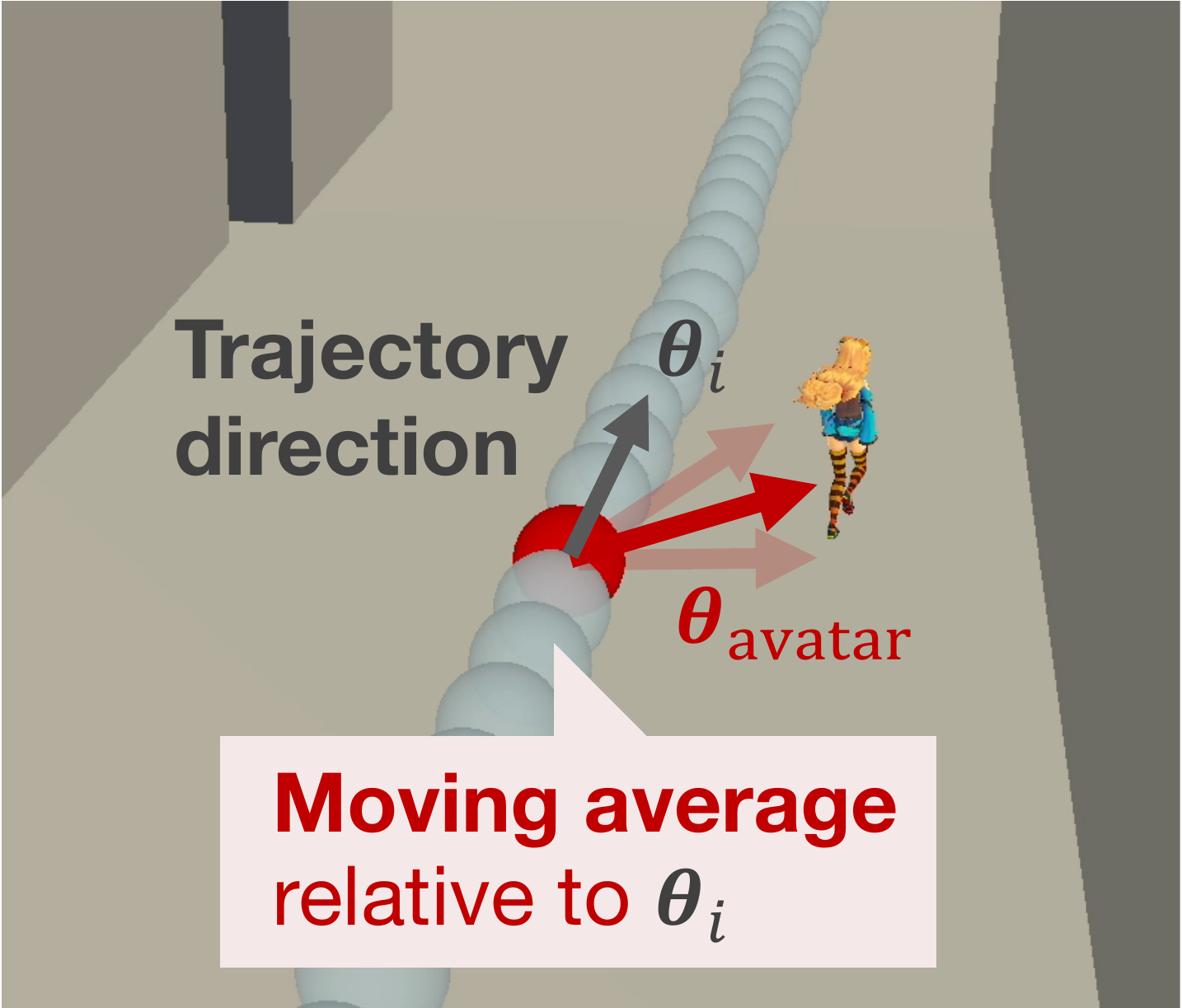}}
    \caption{Overviews of our view-selection strategy. (a) For viewpoint selection, our system calculate the distance $d \in \mathbb{R}$, representing how far the avatar has moved along the trajectory direction. We shift the viewpoint to adjacent frames when the avatar exceeds the threshold distance $\alpha \in \mathbb{R}$. (b) For view orientation selection, we first compute $\bm{\theta}_\mathrm{avatar} \in \mathbb{R}^3$, which denotes the camera orientation that faces the avatar from the current viewpoint. We then apply a temporal moving average to smooth orientation changes relative to the trajectory direction $\bm{\theta}_i \in \mathbb{R}^3$, effectively reducing unwanted oscillations.}
    \label{fig:selection}
\end{figure}

\noindent\emph{Selection of viewpoint:} An overview of the viewpoint selection is shown in Fig.~\ref{fig:selection}~(a). To ensure smooth transitions between viewpoints, we limit the selection of the next viewpoint to adjacent frames, \ie, the $(i-1)$-th or $(i+1)$-th frame when currently at the $i$-th frame.

To determine the next viewpoint, we use the distance the avatar has moved along the forward direction of the trajectory from the current viewpoint. Since the camera's orientation and the trajectory's forward direction are aligned in the walkthrough video, this distance, denoted as $d\in\mathbb{R}$, is calculated as follows:

\begin{equation}
  d = \bm{e}(\bm{\theta}_i) \cdot (\bm{x}_\mathrm{avatar} - \bm{x}_i)
\end{equation}
Here, $\bm{x}_i \in \mathbb{R}^3$ and $\bm{\theta}_i \in \mathbb{R}^3$ denote the position and orientation of the current frame, while $x_\mathrm{avatar}$ denotes the current position of the avatar. Additionally, $\bm{e}(\bm{\theta}_i) \in \mathbb{R}^3$ denotes the unit vector aligned with $\bm{\theta}_i$ and $\cdot$ represents the inner-product of two vectors. Note that $d$ represents the signed distance; a positive $d$ indicates that the avatar has moved forward along the trajectory, while a negative $d$ indicates backward movement.

Based on this distance, we adjust the viewpoint to smoothly follow the avatar's movement while ensuring that the complete avatar remains visible. For this we define the distance that should be kept between the camera and avatar as $\alpha$. Specifically, we shift the viewpoint to the $(i+1)$-th frame when $d > \alpha$ and to the $(i-1)$-th frame when $d < -\alpha$. In our experiments, we empirically set $\alpha$ to 5 meters.
\\

\noindent\emph{Selection of view orientation:} 
An overview of the view orientation selection is shown in Fig.~\ref{fig:selection}~(b). To smoothly track the avatar's movement, we first compute $\bm{\theta}_\mathrm{avatar} \in \mathbb{R}^3$, which represents the camera orientation that faces the avatar from the current viewpoint. However, directly applying this orientation can lead to unwanted oscillations during viewpoint transitions, as the trajectory’s forward direction, denoted as $\bm{\theta}_\mathrm{i} \in \mathbb{R}^3$, may not always align with the avatar’s movement. To reduce these oscillations, we apply a temporal moving average to smooth the orientation changes relative to the trajectory.

First, by converting $\bm{\theta}_i$ and $\bm{\theta}_\mathrm{avatar}$ into quaternions $q_i \in \mathbb{H}$ and $q_\mathrm{avatar} \in \mathbb{H}$, we calculate the quaternion $q'_\mathrm{avatar} \in \mathbb{H}$, which represents orientations of avatar's direction relative to the forward trajectory direction, as follows:
\begin{equation}
    q'_\mathrm{avatar} = q_\mathrm{avatar} * q_i^{-1}
\end{equation}

Next, let $q'_\mathrm{view}(t) \in \mathbb{H}$ represent the relative quaternion of the viewpoint at time $t$ with respect to the forward trajectory direction. To reduce oscillations, we apply a temporal moving average to $q'_\mathrm{view}(t)$ and $q'_\mathrm{view}(t-1)$ using spherical linear interpolation (SLERP). Denoting SLERP between quaternions $q_1 \in \mathbb{H}$ and $q_2 \in \mathbb{H}$, with an interpolation ratio $p \in [0, 1]$ as $\mathtt{slerp}(q_1, q_2, p)$, we calculate $q'_\mathrm{view}(t)$ as follows:
\begin{equation} q'_\mathrm{view}(t) = \mathtt{slerp}(q'_\mathrm{avatar}, q'_\mathrm{view}(t-1), 0.96) \end{equation}
Finally, the resulting view orientation, $q_\mathrm{view}(t) \in \mathbb{H}$, is computed as follows: \begin{equation}
q_\mathrm{view}(t) = q'_\mathrm{view}(t) * q_i
\end{equation}

\section{Experiment \& Result}
In our experiment, we focus on the urban area of Akihabara, Japan, which covers approximately $0.5\,\si{km^{2}}$. We use CityGML data from the PLATEAU project~\cite{PLATEAU}, provided by the Minitsry of Land, Infrastructure, Transport and Tourism of Japan, which includes LOD1 models of buildings and terrain, as well as geospatial data such as building attributes (\eg, address, building height) and flood risk data. Additionally, we use 229 omnidirectional videos (about 130K frames in total) captured along each street within the area. To mitigate camera shake caused by walking, we stabilized the videos using the adaptive subsampling method~\cite{ogawa2017hyperlapse} which downsamples the original 30 FPS footage by a factor of 1/5.\\

\begin{figure}[t]
    \centering
  \subfloat[\label{1a}]{%
       \includegraphics[width=0.32\linewidth]{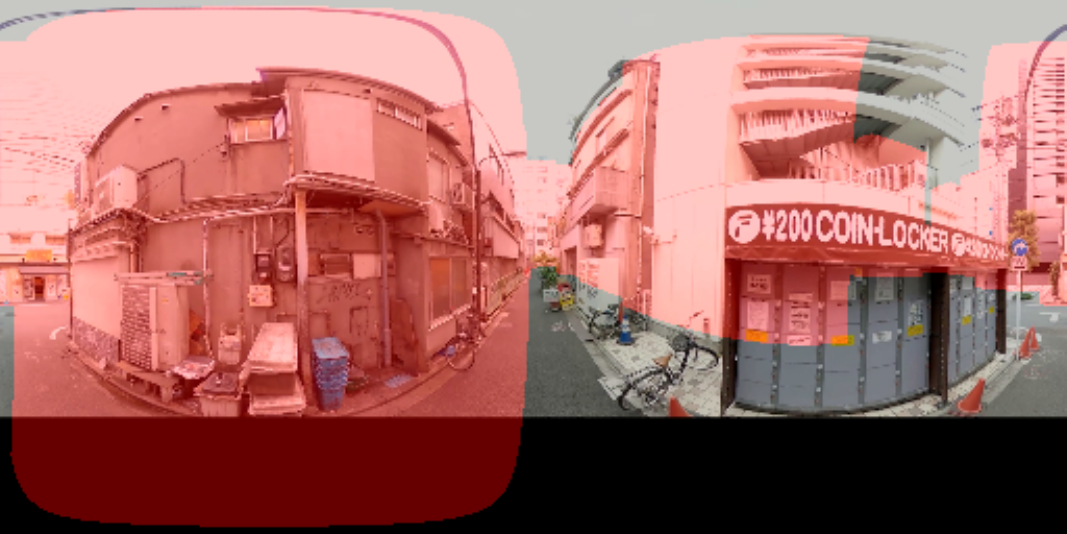}
       \includegraphics[width=0.32\linewidth]{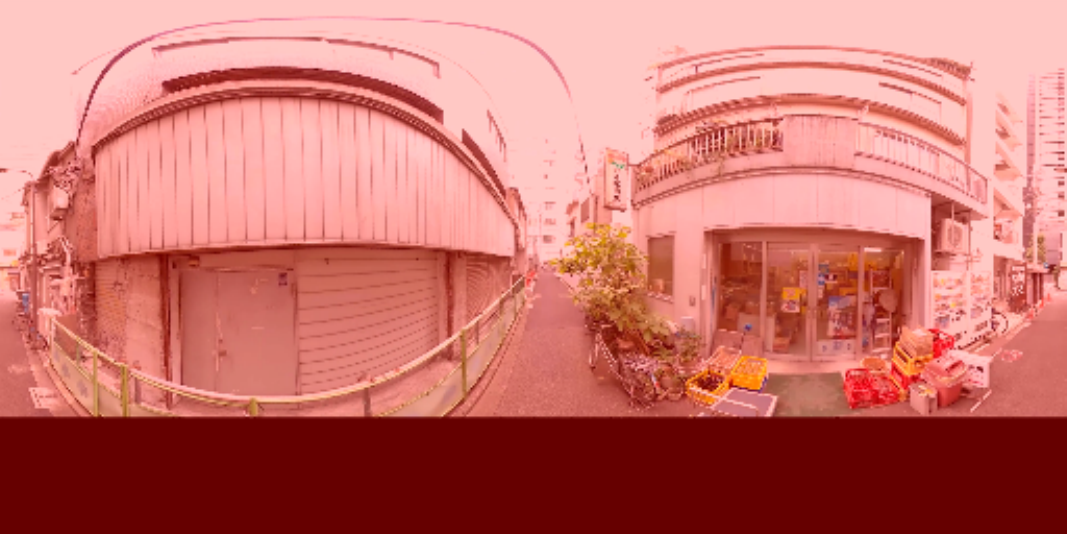}
       \includegraphics[width=0.32\linewidth]{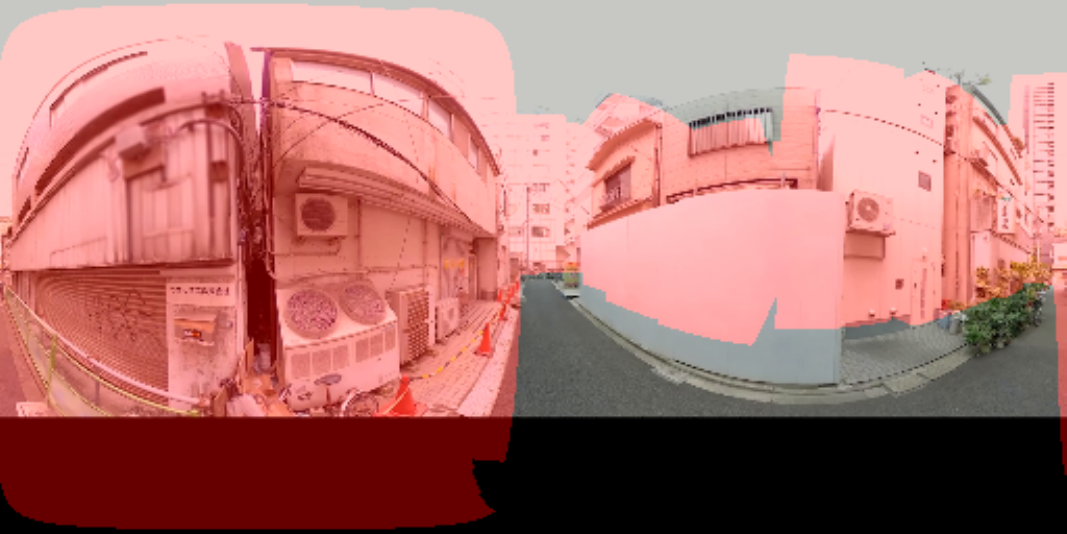}}
    \hfill
  \subfloat[\label{1b}]{%
        \includegraphics[width=0.32\linewidth]{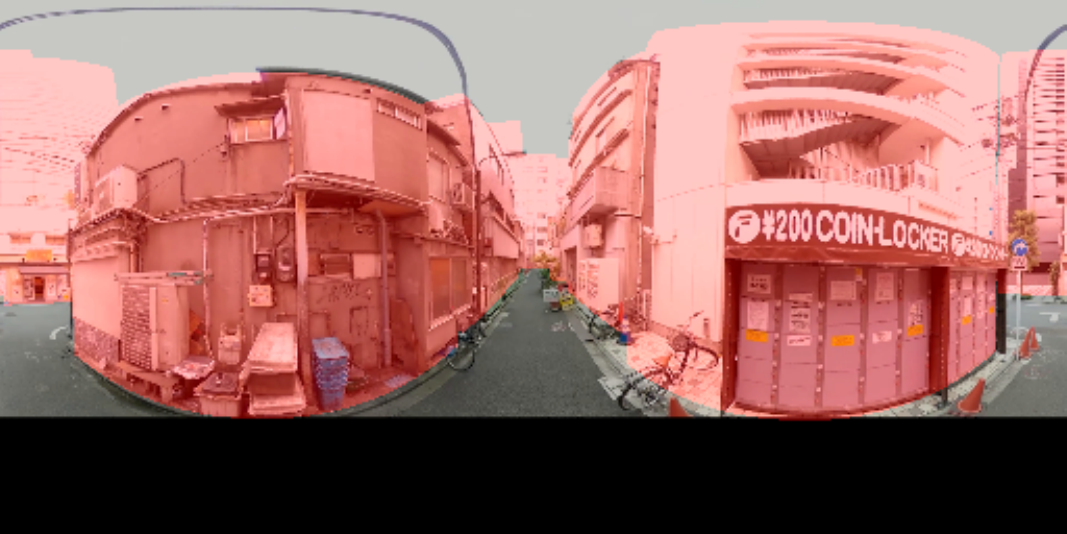}
       \includegraphics[width=0.32\linewidth]{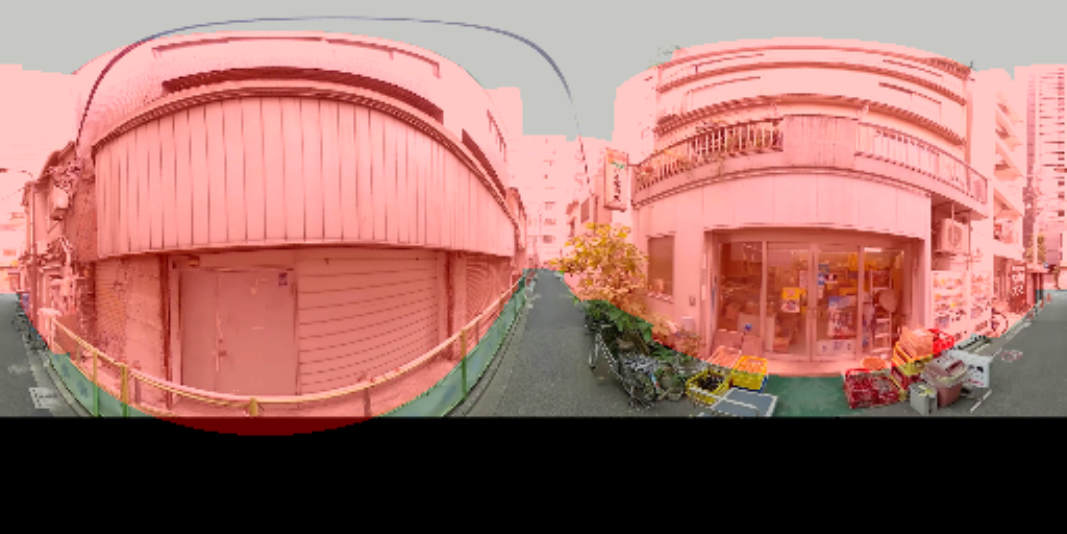}
       \includegraphics[width=0.32\linewidth]{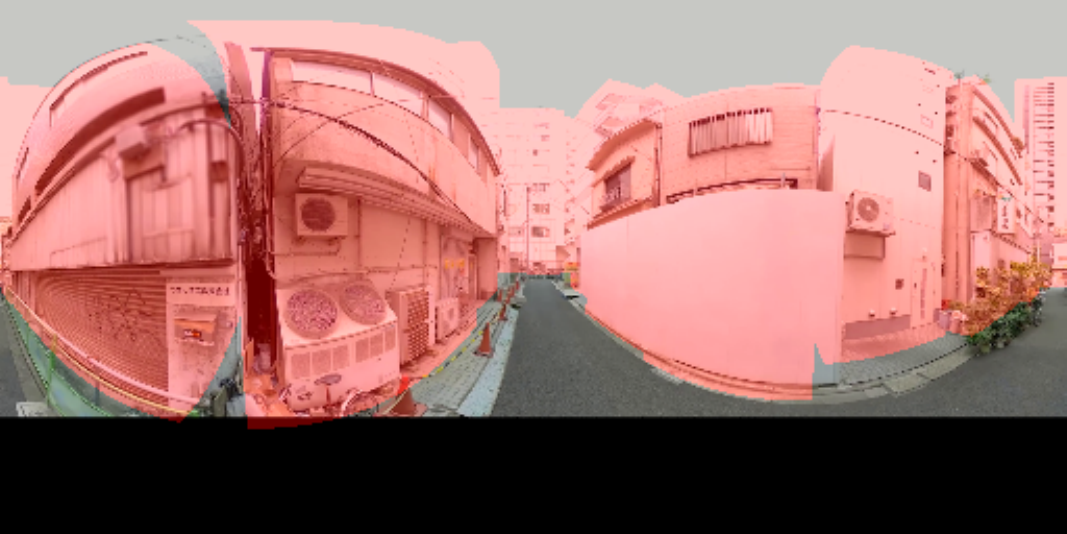}}
  \caption{Comparison of discrepancies between 360° video frames and the CityGML model. (a) before optimization (using initial parameters) and (b) after optimization (using optimized parameters). Red-highlighted areas indicate the projections of the CityGML building models onto the 360° camera view, based on the estimated camera poses for each frame. The optimization process effectively reduces the discrepancies that were visible before optimization.
}
  \label{fig:alignment_qualitative} 
\end{figure}

\noindent\emph{Qualitative result of the alignment:} 
To qualitatively demonstrate the effectiveness of our alignment optimization process, we compare the discrepancies between 360° video frames and the CityGML model before and after optimization, as shown in Fig.~\ref{fig:alignment_qualitative}. The misalignments visible in the pre-optimization view (Fig.~\ref{fig:alignment_qualitative}~(a)) are effectively corrected in the post-optimization view (Fig.~\ref{fig:alignment_qualitative}~(b)), highlighting  the success of our alignment process.
\\

\noindent\emph{Quantitative result of the alignment:} 
Our alignment algorithm uses only sampled frames to evaluate the discrepancies between the video and the CityGML model. To evaluate reduction of discrepancies for frames not included in the sampled set, we conducted the following experiment.

\begin{figure}[t]
    \centering
    \includegraphics[width=.8\linewidth]{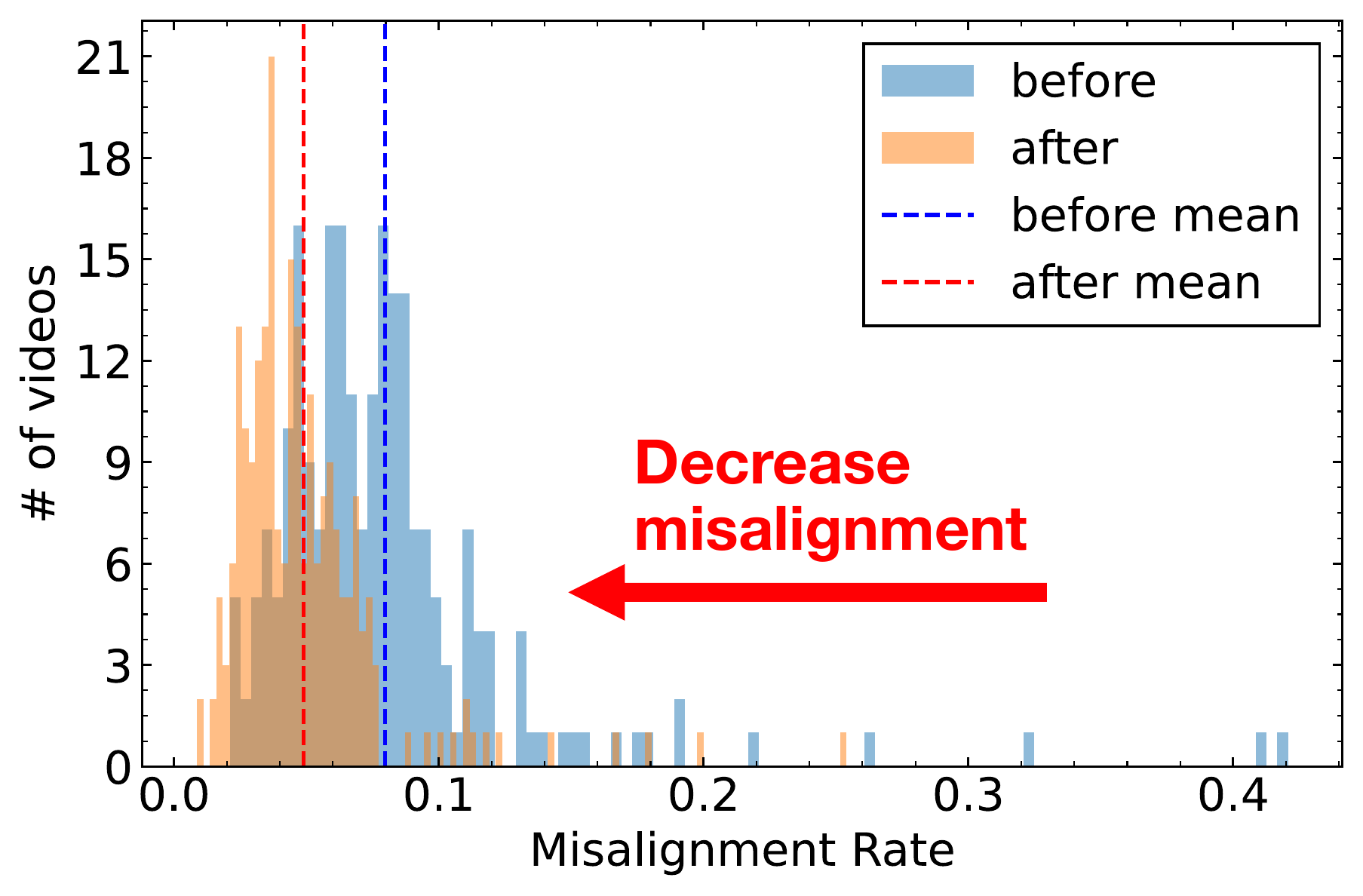}
    \caption{Distribution of Misalignment Rates (number of discrepancy pixels divided by the total number of pixels) calculated across all frames not included the sample set, both before optimization and after optimization for all 229 street videos in the Akihabara area. The post-optimization distribution shows a notable decrease in misalignment rates compared to the pre-optimization distribution, achieving a relative reduction of 39\% in the mean. }
    \label{fig:alignment_quantitative}
\end{figure}

First, we select all frames which are not sampled during the calculation of the objective function, and project the CityGML building models onto the 360° camera view using estimated camera poses, both before and after optimization. Next, for each street video, we calculated the Misalignment Rate by measuring the ratio of the number of discrepant pixels to the total number of pixels across all frames outside the sampled set, both before and after optimization. To ensure accurate comparison, we exclude pixels of regions that are occluded in the video frames (\eg, due to pedestrians, cars, or trees) and only use non-occluded pixels for evaluation. 

Fig.~\ref{fig:alignment_quantitative} shows the distribution of the Misalignment Rate before and after optimization for all 229 street videos in the Akihabara area. Compared to the pre-optimization distribution (mean: 0.0798, standard deviation: 0.0496), the post-optimization distribution (mean: 0.0490, standard deviation: 0.0292) shows a notable decrease in misalignment rates, achieving a relative reduction of 39\% in the mean. This indicates that our optimization effectively aligns even the frames not included in the sampled set.
\\

\section{Application \& User Study}
By incorporating various types of geospatial data, our system is capable of supporting a wide range of urban visualizations in a photorealisic environment. To showcase the practical applications of our system, we present two use cases: flood risk visualization and daylight visualization. Both visualizations were implemented using Unity~\cite{unity} and operate at real-time performance (approximately 100 FPS on a MacBook M2 with 16 GB of memory). In addition, to systematically investigate the advantage of photorealism produced by 360° video on the geospatial data visualization, we conducted a focused user study on flood risk scenearios.

\subsection{Application: flood risk visualization}
\label{sec:flood}

\begin{figure}[t]
    \centering
    \subfloat[\label{1a}]{%
       \includegraphics[width=0.9\linewidth]{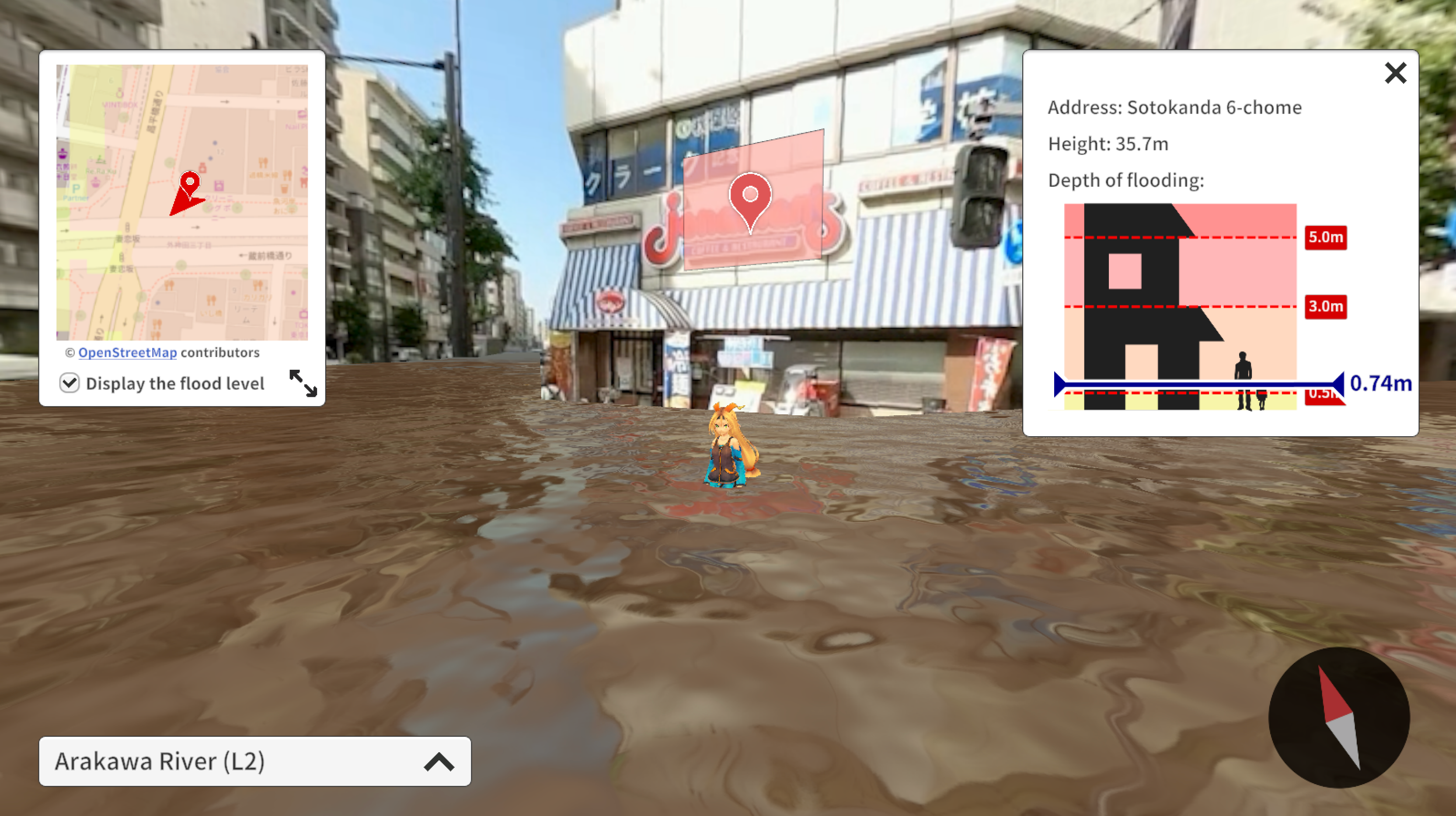}}
\hspace{0.01\linewidth}\\
  \subfloat[\label{1b}]{%
       \includegraphics[width=0.435\linewidth]{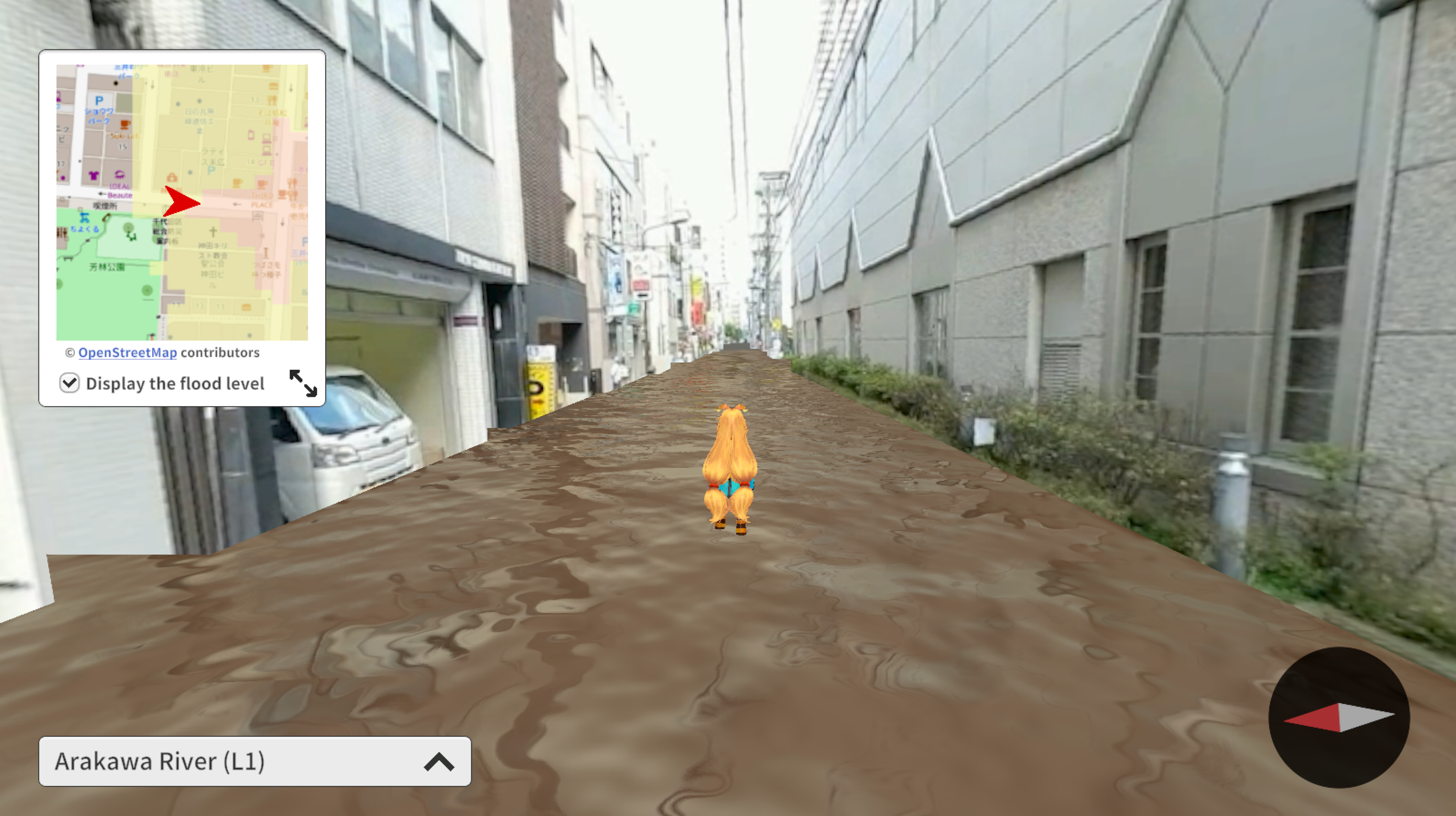}}
\hspace{0.014\linewidth}
  \subfloat[\label{1c}]{%
        \includegraphics[width=0.435\linewidth]{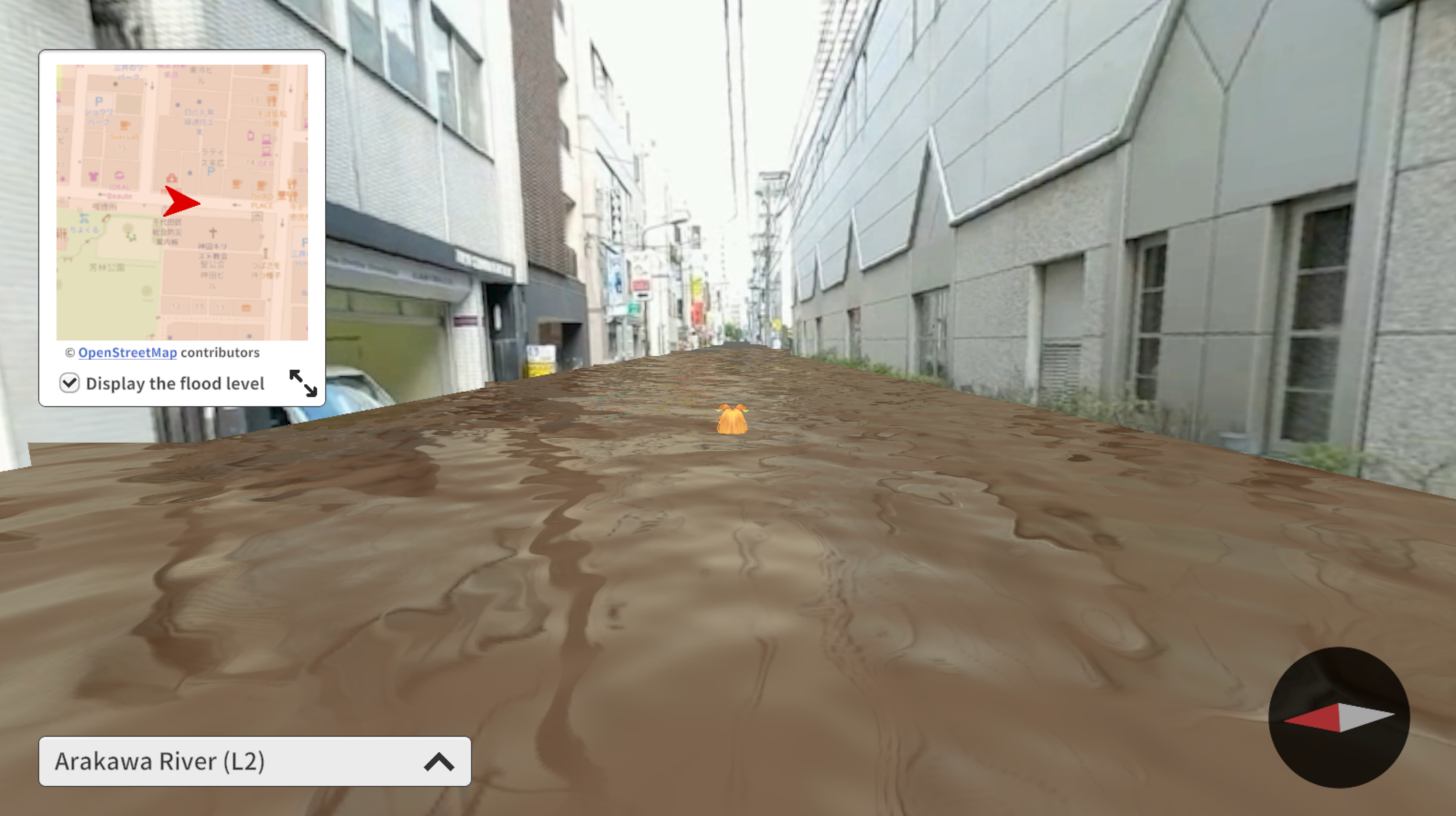}}
  \caption{An example of our flood risk visualization. We use water body geometries representing the predicted flood levels of rivers, as derived from CityGML data~\cite{PLATEAU}. Our visualization allows the user to view flood depth information and compare different water levels. (a) The user can view flood depth information for each building to better understand the associated flood risk. (b) Predicted water levels for Arakawa River (L1: planned scale rainfall). (c) Predicted water levels for Arakawa River (L2: maximum scale rainfall).}
  \label{fig:flood} 
\end{figure}

Fig.~\ref{fig:flood} illustrates an example of our flood risk visualization. In this visualization, we use the water body geometry representing the predicted flood levels of rivers, as derived from CityGML data\cite{PLATEAU}. These geometries, enhanced with a water shader~\cite{AQUAS}, is overlaid onto the CityGML urban 3D model textured by our system. Unlike previous flood visualizations that relied on static texture mapping~\cite{boorboor2024submerse}, our approach directly incorporates 360° video, offering a photorealistic representation of water levels within a real-world context. This enables an intuitive means of conveying flood risks to stakeholders, such as local residents.

In our system, when a user clicks on a building, they can view the flood depth information specific to that building, along with building attributes from the CityGML data~\cite{PLATEAU}, such as address and height (Fig.~\ref{fig:flood}-top). This allows the user to better understand the flood risk for each building. Moreover, the CityGML data~\cite{PLATEAU} includes flood level data for various scenarios, such as different rivers (\eg, Arakawa River or Kanda River) and different rainfall intensities (\eg, L1: planned scale rainfall or L2: assumed maximum scale rainfall). By switching between these scenarios, users can easily compare and understand the differences in predicted flood levels (Fig.~\ref{fig:flood}-bottom).

\subsection{Application: daylight visualization}

\begin{figure}[t]
    \centering
    \subfloat[\label{1a}]{%
        \includegraphics[width=0.45\linewidth]{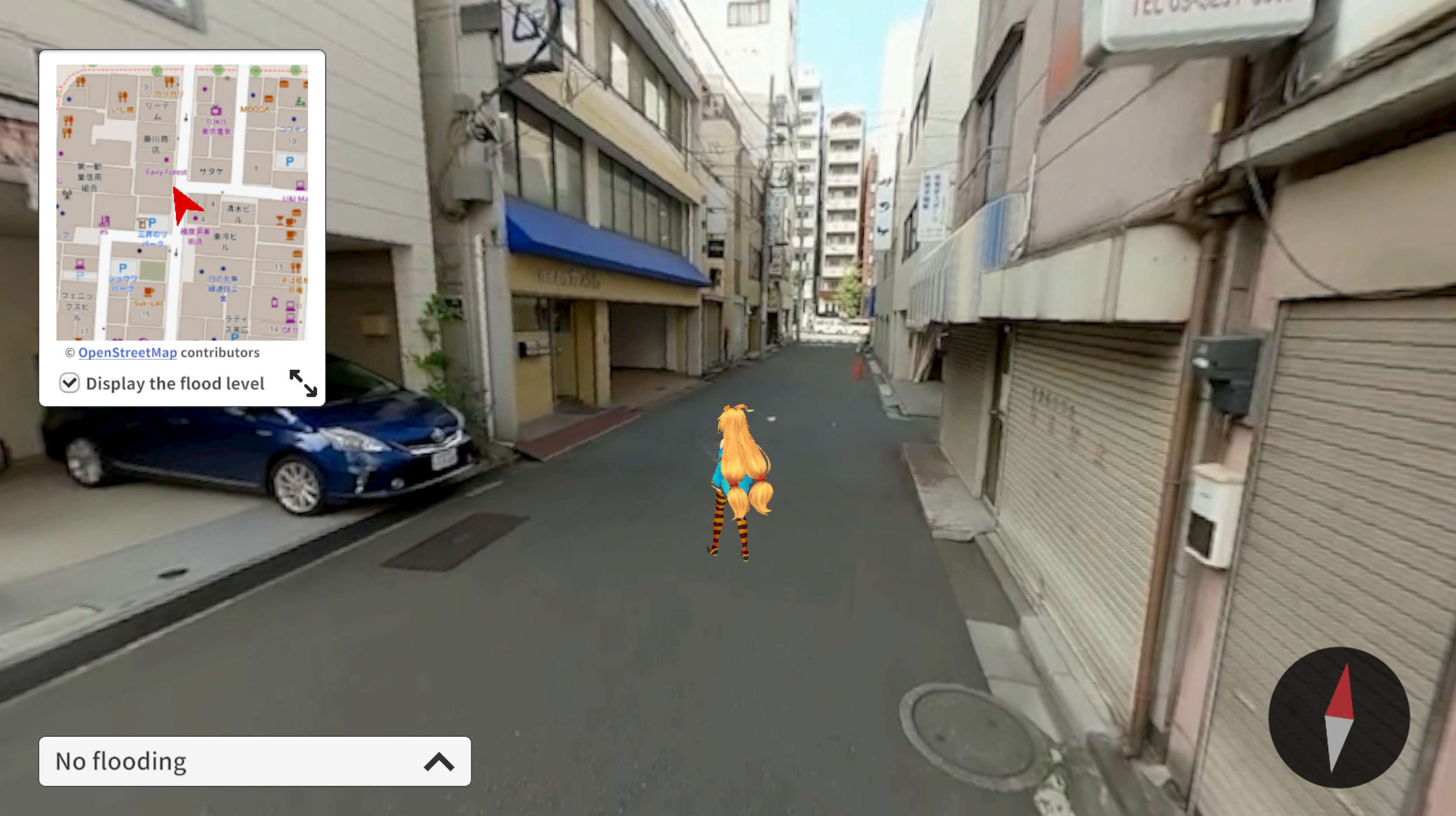}}\\
  \subfloat[\label{1b}]{%
       \includegraphics[width=0.45\linewidth]{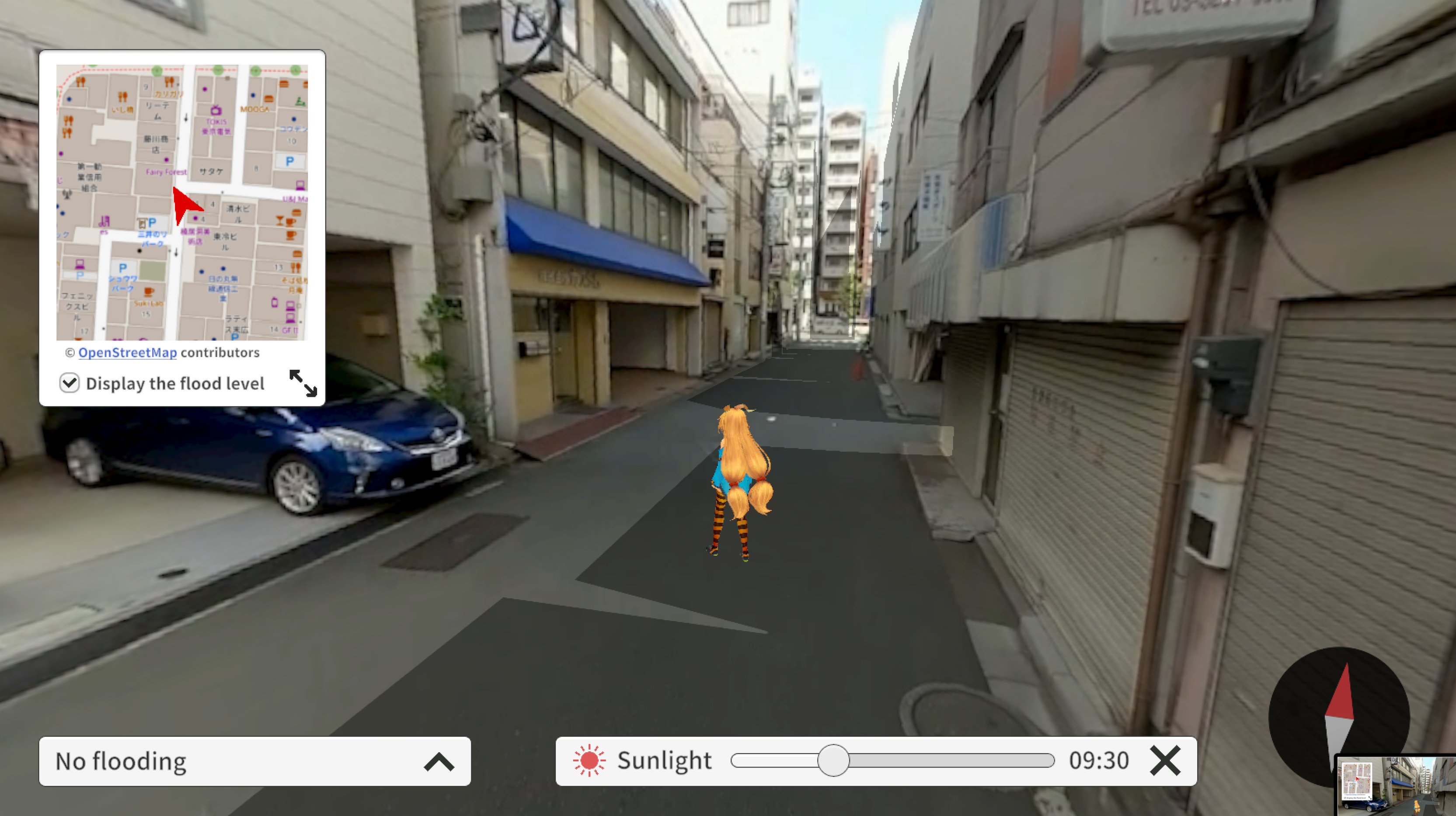}}
\hspace{0.01\linewidth}
  \subfloat[\label{1c}]{%
        \includegraphics[width=0.45\linewidth]{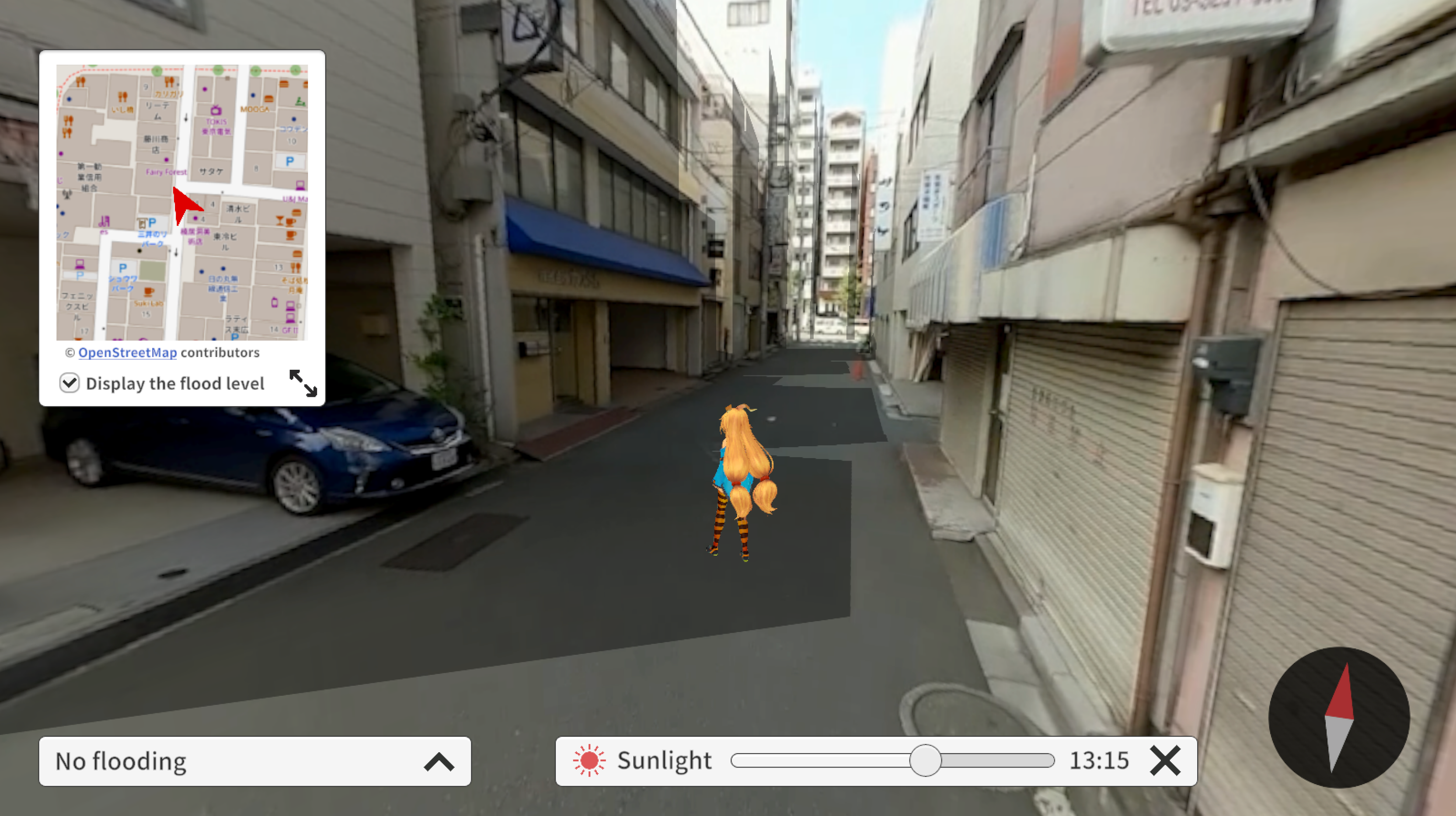}}
  \caption{Examples of our daylight  visualization. 
 (a) Visualization without shadows, (b) Visualization of shadows in the morning,
and (c) Visualization of shadows in the daytime. Our system approximates the sun as a parallel light source and simulates the shadows using the geometry of CityGML. Users can interactively select the time of day to observe and compare shadow patterns at different times.}
  \label{fig:daylight} 
\end{figure}

Fig.~\ref{fig:daylight} illustrates examples of our daylight visualization. In this visualization, the system approximates the sun as a parallel light source and simulates shadows based on the geometry of CityGML data. These simulated shadows are then overlaid onto the CityGML urban 3D model textured by our system. Users can interactively select the time of day to observe and compare shadow patterns at different times.

Despite the existence of various daylight visualization systems~\cite{johnston2001using, miranda2018shadow}, a recent survey~\cite{miranda2024state} concluded that no system offers the exploration of shadow information in urban environments from a pedestrian perspective. Our daylight visualization specifically addresses this gap, and with the rich real-world context provided by 360° video, our system allows users to better understand how shadows impact daily life in residential or commercial areas.

\subsection{User Study on flood risk visualization}
To better understand how photorealism influences users’ interpretation of geospatial data, we conducted a user study on our system, with a particular focus on flood risk visualization. We hypothesize that the high level of photorealism provided by 360° videos enhances users’ sense of presence within the scene, thereby facilitating a more natural and intuitive understanding of the data. To evaluate this hypothesis, we designed the following experiment. The study was approved by our university’s ethics committee. Informed consent was obtained from all participants prior to the experiment.\\

\noindent\emph{Stimuli:}
For our experiment, we used the flood risk visualization scenario of Akihabara introduced in Sec.~\ref{sec:flood}. First, we selected nine streets from this scenario as the targets of our study. For each street, we created three visualization conditions to compare different levels of realism:
\begin{itemize}
    \item \textbf{Non-texture}: Displays the urban 3D model using flat-colored geometry without any surface textures.
    \item \textbf{Static-texture}: Applies predefined texture images to each surface of the urban 3D model.
    \item \textbf{Dynamic-texture}: Our proposed method, which dynamically projects 360° video frames onto the 3D surfaces based on the avatar’s position.
\end{itemize}

\begin{figure}[t]
    \centering
  \subfloat[\label{1a}]{%
       \includegraphics[width=0.31\linewidth]{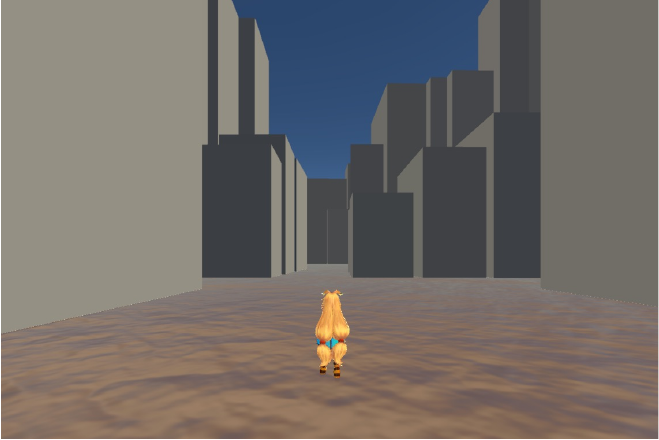}}
\hspace{0.01\linewidth}
  \subfloat[\label{1b}]{%
        \includegraphics[width=0.31\linewidth]{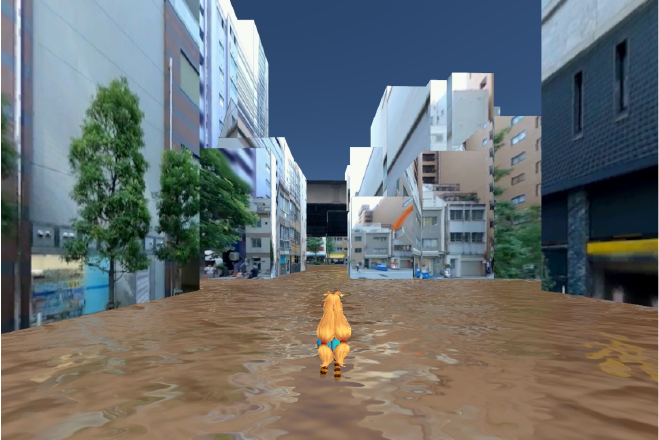}}
\hspace{0.01\linewidth}
  \subfloat[\label{1b}]{%
        \includegraphics[width=0.31\linewidth]{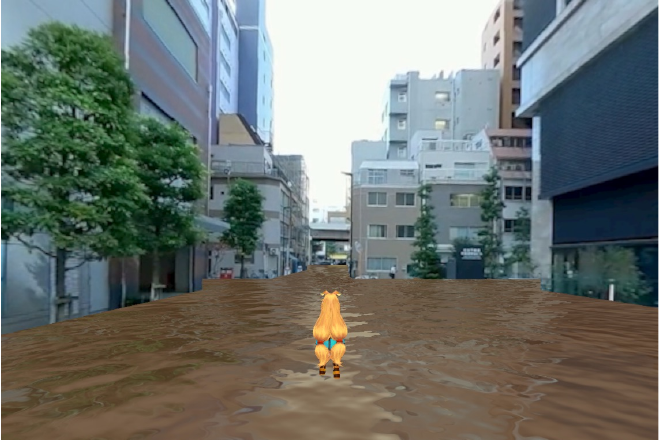}}
  \caption{Example scenes of each condition of our user study. (a) non-texture, (b) static-texture, (c) dynamic-texture (ours).}
  \label{fig:texture_compare_user_study} 
\end{figure}

The static textures were made from the same 360° video used in the dynamic-texture condition, by cropping regions from the video frame spatially closest to each building surface.  Fig.~\ref{fig:texture_compare_user_study} shows example scenes from each condition. Since our system relies on simplified geometric models (\ie, LOD1), static-texture mapping often results in visible misalignments between the geometry and the applied textures. In contrast, the dynamic-texture approach can utilize the real-world imagery directly, which achieves a higher level of photorealism.\\

\noindent\emph{Experiment design:}
We recruited 27 participants (16 male, 11 female; mean age = 28.11, SD = 10.46) to participate in an on-site study. Each participant navigated an avatar through nine streets presented in random order, with each street experienced under one of three assigned conditions. Each condition was presented exactly three times per participant to ensure balanced exposure. Stimulus presentation was counterbalanced throughout the experiment, resulting in an equal number of observations for every condition-location pair.\\

\begin{figure}[t]
    \centering
    \includegraphics[width=0.79\linewidth]{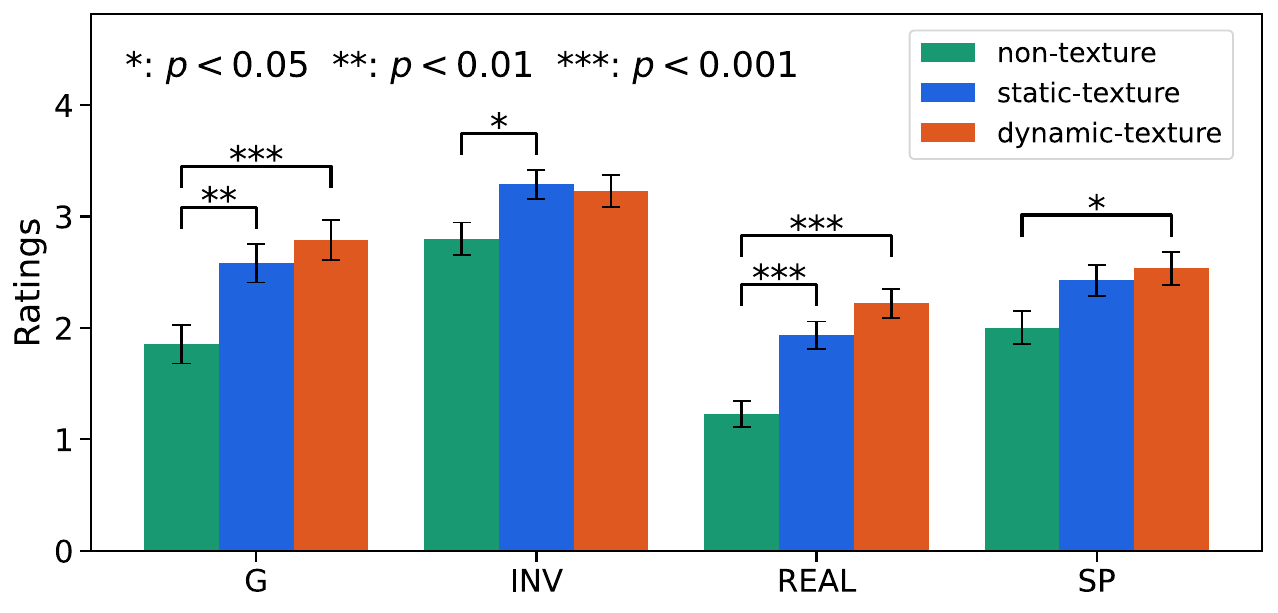}
    \caption{IPQ presence ratings~\cite{schubert2001experience} for each condition, including General Presence (G), Involvement (INV), Experienced Realism (REAL), and Spatial Presence (SP).}
    \label{fig:IPQresult}
\end{figure}

\noindent\emph{Task and measurements:}
For each street, participants navigated an avatar from a starting point to a goal and reported perceived water levels at both ends. After that, they completed the iGroup Presence Questionnaire (IPQ)~\cite{schubert2001experience}, a widely used tool for measuring users’ sense of presence in virtual environments. The IPQ evaluates four subscales: General Presence (G), Involvement (INV), Experienced Realism (REAL), and Spatial Presence (SP), each rated on a seven-point Likert scale (1 = low, 7 = high). Participants concluded the study with a debriefing session, offering qualitative feedback on their experiences across all conditions.\\

\noindent\emph{Results:} 
Fig.~\ref{fig:IPQresult} presents the IPQ scores for each condition. For the General Presence and Experienced Realism measures, both the static-texture and dynamic-texture conditions showed significant improvements compared to the non-texture condition ($p < 0.01^{**}$ for static-texture in General Presence, and $p < 0.001^{***}$ for the other comparisons). This indicates that, compared to the unrealistic non-texture condition, adding textures significantly enhances participants’ sense of immersion and realism. Although there were no statistically significant differences between the static-texture and dynamic-texture conditions, the dynamic-texture condition tended to yield higher scores in General Presence, Experienced Realism, and Spatial Presence.

Interestingly, despite these differences in perceived immersion, the average error in perceived water level was 17.9\,\si{\centi\meter} (SD = 16.7\,\si{\centi\meter}), with no significant differences across conditions: 17.6\,\si{\centi\meter} for non-texture, 16.3\,\si{\centi\meter} for static-texture, and 20.0\,\si{\centi\meter} for dynamic-texture. This is likely due to the fact that 24 of 27 participants, as revealed during debriefing,  used the avatar’s constant height as a primary visual reference for judging water depth, highlighting the avatar’s role as a valuable cue for estimating water levels.

However, despite no significant difference in measured errors, debriefing responses suggested that dynamic-texture condition led to a better subjective understanding of the flood risk data. For instance, several participants remarked, ``I could really feel how high the water would reach'' (P17), or ``It was easier to feel the flood risk'' (P05) under this condition. Overall, 9 participants reported that the dynamic-texture condition facilitated a intuitive and embodied comprehension of the flood risk. These responses indicate that presenting geospatial data through 360° video frames depicting photorealistic environments serves as an effective tool for fostering an intuitive understanding of the data.

\section{Conclusion}
In this paper, we present \emph{360CityGML}, an novel visualization system that effectively integrates 360° walkthrough videos with CityGML model to realize highly realistic and interactive urban visualizations from a pedestrian view.
By directly incorporating 360° video frames, we significantly enhance the realism and utility of LOD1 CityGML model, enabling advanced visualizations of urban phenomena such as flood risks and daylight conditions. We believe our system will broaden the applicability of CityGML data, opening new possibilities for urban data visualization and interpretation.

\bibliographystyle{IEEEtran}
\bibliography{main}

\begin{thebibliography}{10}
\providecommand{\url}[1]{#1}
\csname url@samestyle\endcsname
\providecommand{\newblock}{\relax}
\providecommand{\bibinfo}[2]{#2}
\providecommand{\BIBentrySTDinterwordspacing}{\spaceskip=0pt\relax}
\providecommand{\BIBentryALTinterwordstretchfactor}{4}
\providecommand{\BIBentryALTinterwordspacing}{\spaceskip=\fontdimen2\font plus
\BIBentryALTinterwordstretchfactor\fontdimen3\font minus \fontdimen4\font\relax}
\providecommand{\BIBforeignlanguage}[2]{{%
\expandafter\ifx\csname l@#1\endcsname\relax
\typeout{** WARNING: IEEEtran.bst: No hyphenation pattern has been}%
\typeout{** loaded for the language `#1'. Using the pattern for}%
\typeout{** the default language instead.}%
\else
\language=\csname l@#1\endcsname
\fi
#2}}
\providecommand{\BIBdecl}{\relax}
\BIBdecl

\bibitem{groger2012citygml}
G.~Gr{\"o}ger and L.~Pl{\"u}mer, ``Citygml--interoperable semantic 3d city models,'' \emph{ISPRS P\&RS}, vol.~71, pp. 12--33, 2012.

\bibitem{PLATEAU}
\BIBentryALTinterwordspacing
{MLIT, Japan}, ``Plateau,'' 2020. [Online]. Available: \url{https://www.mlit.go.jp/plateau/}
\BIBentrySTDinterwordspacing

\bibitem{opencitymodel}
\BIBentryALTinterwordspacing
{BuildZero.Org }, ``Open city model,'' 2019. [Online]. Available: \url{https://github.com/opencitymodel/opencitymodel}
\BIBentrySTDinterwordspacing

\bibitem{dollner2006virtual}
J.~D{\"o}llner, T.~H. Kolbe, F.~Liecke, T.~Sgouros, and K.~Teichmann, ``The virtual 3d city model of berlin-managing, integrating, and communicating complex urban information,'' in \emph{UDMS}, 2006.

\bibitem{miranda2024state}
F.~Miranda, T.~Ortner, G.~Moreira, M.~Hosseini, M.~Vuckovic, F.~Biljecki, C.~T. Silva, M.~Lage, and N.~Ferreira, ``The state of the art in visual analytics for 3d urban data,'' in \emph{Computer Graphics Forum}, vol.~43, no.~3.\hskip 1em plus 0.5em minus 0.4em\relax Wiley Online Library, 2024, p. e15112.

\bibitem{yao20183dcitydb}
Z.~Yao, C.~Nagel, F.~Kunde, G.~Hudra, P.~Willkomm, A.~Donaubauer, T.~Adolphi, and T.~H. Kolbe, ``3dcitydb-a 3d geodatabase solution for the management, analysis, and visualization of semantic 3d city models based on citygml,'' \emph{Open geospat. data, softw. stand.}, vol.~3, no.~1, pp. 1--26, 2018.

\bibitem{johnston2001using}
K.~Johnston, J.~M. Ver~Hoef, K.~Krivoruchko, and N.~Lucas, \emph{Using ArcGIS geostatistical analyst}.\hskip 1em plus 0.5em minus 0.4em\relax Esri Redlands, 2001, vol. 380.

\bibitem{cornel2019interactive}
D.~Cornel, A.~Buttinger-Kreuzhuber, A.~Konev, Z.~Horv{\'a}th, M.~Wimmer, R.~Heidrich, and J.~Waser, ``Interactive visualization of flood and heavy rain simulations,'' in \emph{Computer Graphics Forum}, vol.~38, no.~3.\hskip 1em plus 0.5em minus 0.4em\relax Wiley Online Library, 2019, pp. 25--39.

\bibitem{miranda2018shadow}
F.~Miranda, H.~Doraiswamy, M.~Lage, L.~Wilson, M.~Hsieh, and C.~T. Silva, ``Shadow accrual maps: Efficient accumulation of city-scale shadows over time,'' \emph{IEEE TVCG}, vol.~25, no.~3, pp. 1559--1574, 2018.

\bibitem{bartosh2019immersive}
A.~Bartosh and R.~Gu, ``Immersive representation of urban data,'' in \emph{SimAUD}, 2019, pp. 65--68.

\bibitem{zhang2021urbanvr}
C.~Zhang, W.~Zeng, and L.~Liu, ``Urbanvr: An immersive analytics system for context-aware urban design,'' \emph{Comput.Gr.}, vol.~99, pp. 128--138, 2021.

\bibitem{boorboor2024submerse}
S.~Boorboor, Y.~Kim, P.~Hu, J.~M. Moses, B.~A. Colle, and A.~E. Kaufman, ``Submerse: Visualizing storm surge flooding simulations in immersive display ecologies,'' \emph{IEEE TVCG}, vol.~30, no.~9, pp. 6365--6377, 2024.

\bibitem{opendatasets}
\BIBentryALTinterwordspacing
{3D geoinformation research group at Delft University of Technology}, ``Cities/regions around the world with open datasets,'' 2024. [Online]. Available: \url{https://3d.bk.tudelft.nl/opendata/opencities/}
\BIBentrySTDinterwordspacing

\bibitem{sugimoto2020urban}
N.~Sugimoto, T.~Okubo, and K.~Aizawa, ``Urban movie map for walkers: Route view synthesis using 360 videos,'' in \emph{ICMR}, 2020, pp. 502--508.

\bibitem{sugimoto2020building}
N.~Sugimoto, Y.~Ebine, and K.~Aizawa, ``Building movie map-a tool for exploring areas in a city-and its evaluations,'' in \emph{ACMMM}, 2020, pp. 3330--3338.

\bibitem{takenawa2023360rvw}
M.~Takenawa, N.~Sugimoto, L.~W{\"o}hler, S.~Ikehata, and K.~Aizawa, ``360rvw: Fusing real 360° videos and interactive virtual worlds,'' in \emph{ACMMM}, 2023, pp. 9379--9381.

\bibitem{takenawa2025building}
------, ``Building and evaluating a realistic virtual world for large scale urban exploration from 360° videos,'' \emph{arXiv preprint arXiv:2510.11447}, 2025.

\bibitem{taneja2012registration}
A.~Taneja, L.~Ballan, and M.~Pollefeys, ``Registration of spherical panoramic images with cadastral 3d models,'' in \emph{3DIMPVT}.\hskip 1em plus 0.5em minus 0.4em\relax IEEE, 2012, pp. 479--486.

\bibitem{sumikura2019openvslam}
S.~Sumikura, M.~Shibuya, and K.~Sakurada, ``Openvslam: A versatile visual slam framework,'' in \emph{ACMMM}, 2019, pp. 2292--2295.

\bibitem{blender}
\BIBentryALTinterwordspacing
{The Blender Foundation}, ``{Blender},'' 2024. [Online]. Available: \url{https://www.blender.org/}
\BIBentrySTDinterwordspacing

\bibitem{bredif2013image}
M.~Br{\'e}dif, ``Image-based rendering of lod1 3d city models for traffic-augmented immersive street-view navigation,'' \emph{ISPRS Annals}, vol.~2, pp. 7--11, 2013.

\bibitem{bredif2014projective}
------, ``Projective texturing uncertain geometry: silhouette-aware box-filtered blending using integral radial images,'' \emph{ISPRS Annals}, vol.~2, pp. 17--23, 2014.

\bibitem{du2019geollery}
R.~Du, D.~Li, and A.~Varshney, ``Geollery: A mixed reality social media platform,'' in \emph{CHI}, 2019, pp. 1--13.

\bibitem{park2021instant}
J.~Park, I.-B. Jeon, S.-E. Yoon, and W.~Woo, ``Instant panoramic texture mapping with semantic object matching for large-scale urban scene reproduction,'' \emph{IEEE TVCG}, vol.~27, no.~5, pp. 2746--2756, 2021.

\bibitem{mildenhall2020nerf}
B.~Mildenhall, P.~P. Srinivasan, M.~Tancik, J.~T. Barron, R.~Ramamoorthi, and R.~Ng, ``Nerf: Representing scenes as neural radiance fields for view synthesis,'' in \emph{ECCV}, 2020, pp. 405--421.

\bibitem{kerbl3Dgaussians}
B.~Kerbl, G.~Kopanas, T.~Leimk{\"u}hler, and G.~Drettakis, ``3d gaussian splatting for real-time radiance field rendering,'' \emph{ACM trans. graph.}, vol.~42, no.~4, July 2023.

\bibitem{liu2024citygaussian}
Y.~Liu, C.~Luo, L.~Fan, N.~Wang, J.~Peng, and Z.~Zhang, ``Citygaussian: Real-time high-quality large-scale scene rendering with gaussians,'' in \emph{ECCV}, 2024, pp. 265--282.

\bibitem{otonari2024entity}
T.~Otonari, S.~Ikehata, and K.~Aizawa, ``Entity-nerf: Detecting and removing moving entities in urban scenes,'' in \emph{CVPR}, 2024, pp. 20\,892--20\,901.

\bibitem{chalmers2024avatar360}
A.~Chalmers, F.~Zaman, and T.~Rhee, ``Avatar360: Emulating 6-dof perception in 360° panoramas through avatar-assisted navigation,'' in \emph{IEEE VR}, 2024, pp. 630--638.

\bibitem{cheng2022masked}
B.~Cheng, I.~Misra, A.~G. Schwing, A.~Kirillov, and R.~Girdhar, ``Masked-attention mask transformer for universal image segmentation,'' in \emph{CVPR}, 2022, pp. 1290--1299.

\bibitem{neuhold2017mapillary}
G.~Neuhold, T.~Ollmann, S.~Rota~Bulo, and P.~Kontschieder, ``The mapillary vistas dataset for semantic understanding of street scenes,'' in \emph{ICCV}, 2017, pp. 4990--4999.

\bibitem{hansen2001completely}
N.~Hansen and A.~Ostermeier, ``Completely derandomized self-adaptation in evolution strategies,'' \emph{Evol. Comput.}, vol.~9, no.~2, pp. 159--195, 2001.

\bibitem{ogawa2017hyperlapse}
M.~Ogawa, T.~Yamasaki, and K.~Aizawa, ``Hyperlapse generation of omnidirectional videos by adaptive sampling based on 3d camera positions,'' in \emph{ICIP}.\hskip 1em plus 0.5em minus 0.4em\relax IEEE, 2017, pp. 2124--2128.

\bibitem{unity}
\BIBentryALTinterwordspacing
{Unity Technologies}, ``{Unity},'' 2024. [Online]. Available: \url{https://unity.com}
\BIBentrySTDinterwordspacing

\bibitem{AQUAS}
\BIBentryALTinterwordspacing
Dogmatic, ``Aquas lite - built-in render pipeline,'' 2019. [Online]. Available: \url{https://assetstore.unity.com/packages/vfx/shaders/aquas-lite-built-in-render-pipeline-53519}
\BIBentrySTDinterwordspacing

\bibitem{schubert2001experience}
T.~Schubert, F.~Friedmann, and H.~Regenbrecht, ``The experience of presence: Factor analytic insights,'' \emph{Presence (Camb. Mass.)}, vol.~10, no.~3, pp. 266--281, 2001.

\end{thebibliography}

\vfill

\end{document}